\documentclass[preprint,showpacs,preprintnumbers,amsmath,amssymb,superscriptaddress,prb]{revtex4-2}
\usepackage{graphicx}
\usepackage[utf8]{inputenc}
\usepackage{amssymb}
\usepackage{color}
\usepackage{units}
\usepackage{float}
\usepackage{amsmath}
\usepackage{hyperref}
\usepackage{epsfig}
\usepackage{bm}
\usepackage{ulem}
\usepackage{hyphenat}
\usepackage{makecell}
\usepackage{color,soul}

\begin{document}
\title{Anomalous Nernst effect in Mn$_3$NiN thin films}
\author{Sebastian Beckert}
\affiliation{Institut f{\"u}r Festk{\"o}rper- und Materialphysik, Technische Universit{\"a}t Dresden, 01062 Dresden, Germany}
\author{João Godinho}
\affiliation{Institute of Physics ASCR, v.v.i., Cukrovarnick\'a 10, 162 53, Prague, Czech Republic}
\affiliation{Faculty of Mathematics and Physics, Charles University, Ke Karlovu 3, Prague, Czech Republic}
\author{Freya Johnson}
\affiliation{Department of Physics, Blackett Laboratory, Imperial College London, London SW7 2AZ, United Kingdom}
\author{Jozef Kimák}
\affiliation{Faculty of Mathematics and Physics, Charles University, Ke Karlovu 3, Prague, Czech Republic}
\author{Eva Schmoranzerová}
\affiliation{Faculty of Mathematics and Physics, Charles University, Ke Karlovu 3, Prague, Czech Republic}
\author{Jan Zemen}
\affiliation{Faculty of Electrical Engineering, Czech Technical University in Prague, Technická 2, Prague, Czech Republic}
\author{Zbyněk Šobáň}
\affiliation{Institute of Physics ASCR, v.v.i., Cukrovarnick\'a 10, 162 53, Prague, Czech Republic}
\author{Kamil Olejník}
\affiliation{Institute of Physics ASCR, v.v.i., Cukrovarnick\'a 10, 162 53, Prague, Czech Republic}
\author{Jakub Železný}
\affiliation{Institute of Physics ASCR, v.v.i., Cukrovarnick\'a 10, 162 53, Prague, Czech Republic}
\author{Joerg Wunderlich}
\affiliation{Institute of Experimental and Applied Physics, University of Regensburg, Universit{\"a}tsstraße 31, 93051 Regensburg, Germany}
\author{Petr Němec}
\affiliation{Faculty of Mathematics and Physics, Charles University, Ke Karlovu 3, Prague, Czech Republic}
\author{Dominik Kriegner}
\affiliation{Institut f{\"u}r Festk{\"o}rper- und Materialphysik, Technische Universit{\"a}t Dresden, 01062 Dresden, Germany}
\affiliation{Institute of Physics ASCR, v.v.i., Cukrovarnick\'a 10, 162 53, Prague, Czech Republic}
\author{Andy Thomas}
\affiliation{Institut f{\"u}r Festk{\"o}rper- und Materialphysik, Technische Universit{\"a}t Dresden, 01062 Dresden, Germany}
\affiliation{Leibniz Institute of Solid State and Materials Science (IFW Dresden), Helmholtzstr.\ 20, 01069 Dresden, Germany}
\author{Sebastian T. B. Goennenwein}
\affiliation{Department of Physics, University of Konstanz, 78457 Konstanz, Germany}
\author{Lesley F Cohen}
\affiliation{Department of Physics, Blackett Laboratory, Imperial College London, London SW7 2AZ, United Kingdom}
\author{Helena Reichlová}
\affiliation{Institut f{\"u}r Festk{\"o}rper- und Materialphysik, Technische Universit{\"a}t Dresden, 01062 Dresden, Germany}
\affiliation{Institute of Physics ASCR, v.v.i., Cukrovarnick\'a 10, 162 53, Prague, Czech Republic}

\date{November 25, 2022}

\begin{abstract}


The observation of a sizable anomalous Hall effect in magnetic materials with vanishing magnetization has renewed interest in understanding and engineering this phenomenon. Antiferromagnetic antiperovskites are one of emerging material classes that exhibit a variety of interesting properties owing to a complex electronic band structure and magnetic ordering. Reports on the anomalous Nernst effect and its magnitude in this class of materials are, however, very limited. This scarcity may be partly due to the experimental difficulty of reliably quantifying the anomalous Nernst coefficient. Here, we report experiments on the anomalous Nernst effect in antiferromagnetic antiperovskite Mn$_3$NiN thin films. Measurement of both the anomalous Hall and Nernst effects using the same sample and measurement geometry makes it possible to directly compare these two effects and quantify the anomalous Nernst coefficient and conductivity in Mn$_3$NiN. We carefully evaluate the spatial distribution of the thermal gradient in the sample and use finite element modeling to corroborate our experimental results. 

\end{abstract}

\maketitle

\section{Introduction}
In magnetically ordered material, a thermal gradient can induce a voltage perpendicular to both the thermal gradient direction and the magnetic order vector. This thermally generated transverse voltage is referred to as the anomalous Nernst effect (ANE) and can be understood as a thermal counterpart of the anomalous Hall effect (AHE). In recent years, the ANE has attracted increasing attention because, similar to the AHE, the various underlying mechanisms have been disentangled \cite{ding2019}. A key step was the development of the Berry phase concept, which links the intrinsic AHE and ANE to the integration of the Berry curvature over the first Brillouin zone \cite{chen2014}. When the combined time reversal with inversion symmetry and combined time reversal with translation symmetry operations are broken and the corresponding integral is nonzero, intrinsic contributions to the ANE and AHE can exist. However, the breaking of these combined symmetries  does not imply that the spins are ferromagnetically ordered. As a consequence, the ANE was observed in materials that were previously thought not to allow this phenomenon because of a vanishing net magnetization \cite{ikhlas2017}. Recent progress also showed a path to increase the ANE response, leading to the observation of very large ANE coefficients in emerging materials such as Fe$_3$Sn, UCo$_{0.8}$Ru$_{0.2}$Al and YbMnBi$_2$\cite{chen2022,asaba2021,pan2022} by engineering the electronic band structure and the position of the Fermi level. 

Despite the major advances that have been made in recent years, the ANE is not yet fully understood. The field dependence of the ANE in comparison to that of the AHE and the expected anisotropy in nontrivial materials are still topics under active discussion. \cite{zhou2020a}. The ANE is dominated by the states in the proximity of the Fermi level. Therefore, in materials with a complex band structure the ANE can serve as a more sensitive probe of changes of the Fermi energy $E_\mathrm{F}$, than the AHE, which is proportional to the integral of the Berry curvature over all occupied states \cite{wuttke2019,noky2018}. Another advantage of the ANE is related to the versatility of the thermal gradient. In thin antiferromagnetic films, the crystal and spin orientation are typically determined by epitaxial constraints and cannot be easily modified, therefore, measuring the AHE along different crystal directions can be challenging. In particular, measuring the AHE response for a current perpendicular to the thin-film plane requires tedious lithography and 3D microfabrication \cite{sun1996}. By contrast, applying an out-of-plane thermal gradient using a  focused laser beam is comparatively simple \cite{reichlova2019,janda2020,johnson2022}. However, quantifying the thermal gradient in this measurement geometry is very challenging. Over the last few years, the ANE has also been discussed in the context of applications such as heat flux sensors \cite{zhou2020}. Here, again, the antiferromagnetic ANE would be beneficial owning to the absence of magnetic stray fields. Despite these benefits, the ANE is far less studied than the AHE because control and quantification of an applied thermal gradient is especially challenging and not as straightforward as applying and measuring an electric current, as is used to determine the AHE. 

The promising family of antiperovskites manganese nitrides Mn$_3$XN, where X represents Ni, Sn, or Ga, is actively studied \cite{nan2020,gurung2020,florez-gomez2022}. These materials have been reported to exhibit the AHE \cite{boldrin2019,you2020,hajiri2019} as a consequence of the symmetries broken by the non-collinear magnetic order and the resulting Berry curvature \cite{zhou2020a}. Interestingly, strain control of the AHE was demonstrated \cite{boldrin2018, boldrin2019a,zemen2017,johnson2021}, which is an important step toward strain controlled switching of the AHE response, as was recently reported Mn$_3$Sn \cite{ikhlas2022}. At the same time, very limited reports on the ANE in manganese nitrides are available: the ANE was very recently reported in Mn$_3$SnN \cite{you2022} while applying an out-of-plane thermal gradient. A laser induced thermal gradient was also used to visualize the magnetic structure of Mn$_3$NiN \cite{johnson2022}. However, a systematic evaluation and quantification of the ANE and AHE in a single sample with a well defined thermal gradient has yet to be performed. 

In this paper, we report a systematic study of the anomalous Nernst effect in antiferromagnetic thin-film Mn$_3$NiN. In this study, the ANE is generated by an in-plane thermal gradient, paying special attention to the spatial distribution of the thermal gradient, which we compare to numerical simulations based on the finite element method (FEM) as implemented in COMSOL Multiphysics \cite{comsol}. We measure an anomalous Nernst coefficient of $\unitfrac[0.0382] {\mu V}{K}$ at temperatures between \unit[150]{K} and \unit[190]{K}, which is comparable to the reported Nernst coefficient of the closely related Mn$_3$SnN \cite{you2022}. We propose that Mn$_3$NiN is not only a fundamentally interesting compound, but also an ideal model system, because its non-collinear magnetic structure in the (111)-plane enables the generation of an ANE by both in- and out-of-plane thermal gradients in (001)-oriented thin films \cite{johnson2022}. Moreover, the position of the Fermi energy can be tuned by strain induced from various substrates \cite{boldrin2019a}.

\section{Sample fabrication}
\begin{figure}[H]
	\centering
	\includegraphics[width=\textwidth]{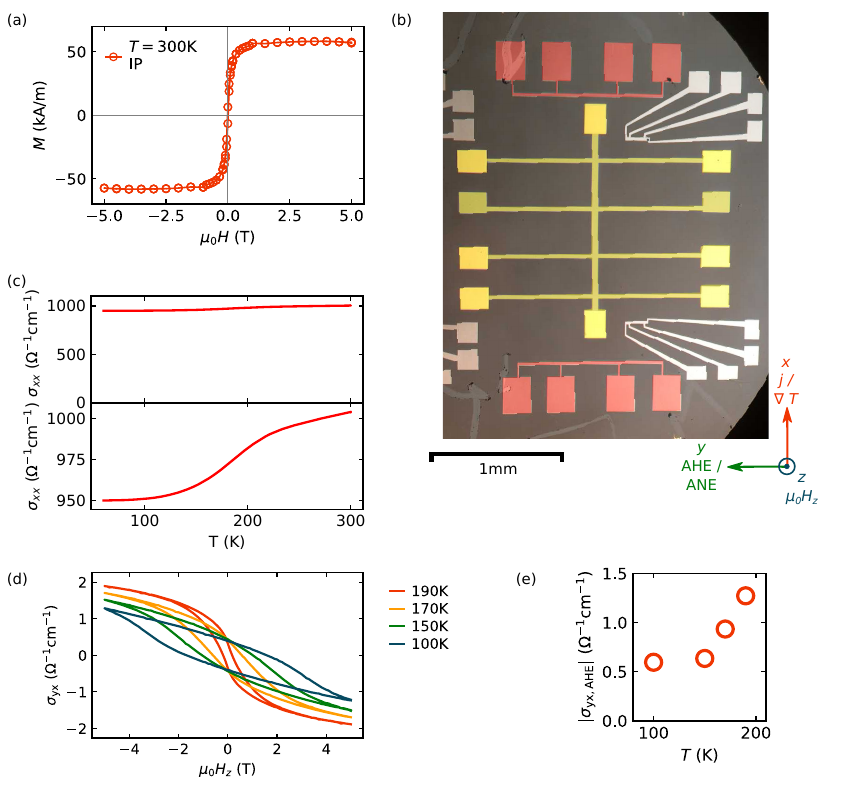}
	\caption{\textbf{Characterization of the Mn$_3$NiN films.} (a) Magnetization loops measured with a magnetic field in the sample plane at 300~K. (b) False-color microscopy image of the Mn$_3$NiN Hall bar (yellow) with platinum heaters (red) and thermometers (white). 
	(c) Temperature dependency of the longitudinal conductivity. 
	(d) Magnetic-field dependence of the Hall conductivity at different temperatures. (e) Temperature dependency of the anomalous Hall conductivity.\label{fig:charac}}
\end{figure}

The Mn$_3$NiN film with a thickness of \unit[50]{nm} used in this study was grown by pulsed laser deposition on a single crystal (001)-oriented SrTiO$_3$ substrate at a temperature of $\unit[400]{^\circ C}$. The film has a [001] film normal and a Néel temperature of $T_\mathrm{N}=\unit[230]{K}$ \cite{johnson2022}. The magnetic properties of the film are in agreement with those found in previous studies \cite{boldrin2019a} (see Fig. \ref{fig:charac}(a)).

The film was patterned into Hall bars with a width of $\unit[45]{\mu m}$ and transversal contacts with a length of $\unit[1800]{\mu m}$ using electron beam lithography and a wet etching process with diluted ferric chloride. After the film was etched, \unit[50]{nm}-thick platinum heaters and thermometers were fabricated by Pt sputtering using a lift-off process. The same sample geometry was previously used to measure the ANE in ferromagnetic Co$_2$MnGa thin films \cite{park2020}. A false-color microscopy image of the resulting device is shown in Fig. \ref{fig:charac}(b).

The sample layout with on-chip heaters and thermometers enables direct measurement and comparison of the AHE and ANE for the same device. Owing to the sample geometry, the in-plane temperature gradient generated by Joule heating of the platinum heater is considerably smaller than the out-of-plane thermal gradient generated by a platinum heater on top of the thin film, as measured by You et al. on Mn$_3$SnN \cite{you2022}. Thus, we measured smaller Nernst voltages in this study. However, this geometry has the advantage of enabling the temperature  to be directly measured at different positions in our sample, for use in evaluating the local temperature gradients, Nernst conductivities, and Nernst coefficients with higher accuracy.

\section{Hall measurements}
Fig \ref{fig:charac}(c) shows the temperature-dependent longitudinal conductivity $\sigma_{xx}$, which suggests a semimetallic character of the material that is consistent with the results of previous studies. The anomalous Hall effect was measured by applying a magnetic field perpendicular to the sample plane (see coordinate system depicted in Fig. \ref{fig:charac}(b)). The Hall resistivity $\rho_{yx}$ was calculated as
\begin{equation}
    \rho_{yx}=\frac{V_y}{I_x}t
\end{equation}
where the transversal voltage is denoted as $V_y$, the sourced current is denoted as $I_x$, and the film thickness is denoted $t$.
We used the Hall resistivity to calculate the Hall conductivity as
\begin{equation}
    \sigma_{yx}=-\frac{\rho_{yx}}{\rho_{xx}^2+\rho_{yx}^2}
\end{equation}
where the longitudinal resistivity is denoted by $\rho_{xx}$ \cite{pu2008}.  The ordinary Hall contribution was subtracted via linear fitting to the Hall voltage at high fields to extract the anomalous contribution. 
The total Hall conductivity $\sigma_{{yx}}=\sigma_{{yx}\mathrm{,OHE}}+\sigma_{{yx}\mathrm{,AHE}}$ measured at various temperatures is shown in Fig. \ref{fig:charac}(d), and the temperature dependency of the anomalous Hall conductivity at saturated magnetization is shown in Fig. \ref{fig:charac}(e). Upon cooling, the anomalous Hall conductivity decreases and the coercive field increases, in agreement with previous observations of Mn$_3$NiN films prepared on STO substrates \cite{Boldrin_AHE_Mn3NiN_Arxiv}\cite{boldrin2019}. Assuming that the Mott rule is valid \cite{pu2008} \cite{Smrcka_1977}, the AHE provides an indication of the field dependence of the ANE signal.

\section{Nernst measurements and analysis of temperature gradient}
\begin{figure}[H]
	\centering
	\includegraphics[width=\textwidth]{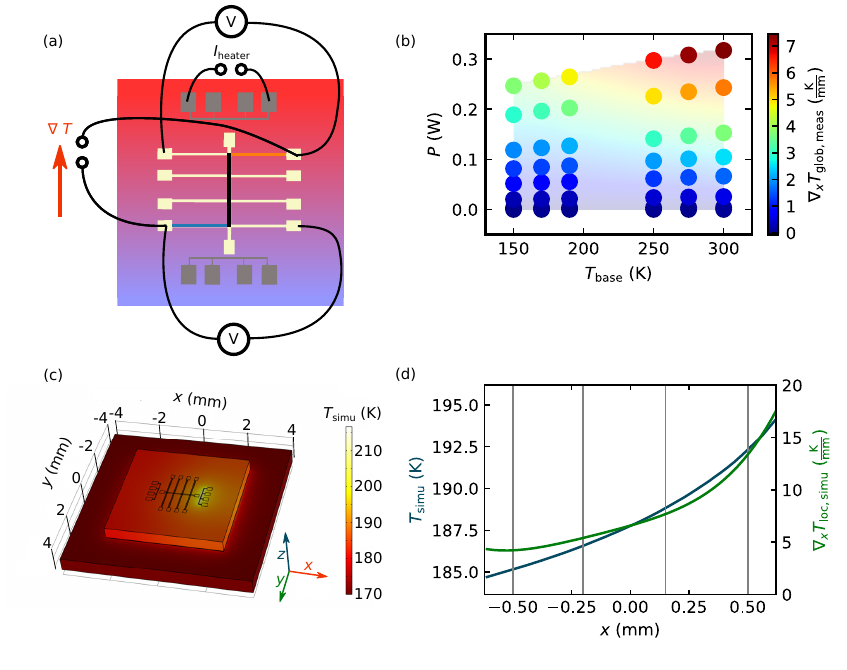}
	\caption{\textbf{Calibration of the local temperature gradients.} (a) Schematic of the temperature gradient applied to the sample, with an example of the measurement geometry used to evaluate the global temperature gradient with the sample arms as resistive thermometers. The sections of the Hall bar arm highlighted in orange and blue denote the temperatures at the hot and cold sides of the bar, respectively. (b) Measured values of the thermal gradient (points) and the corresponding linear interpolation (colored area) as a function of the heater power and the sample base temperature $T_\mathrm{base}$. (c) FEM simulated temperature profile over the sample for $T_\mathrm{base}=\unit[170]{K}$. (d) FEM simulated temperature profile and temperature gradients along the Hall bar for $T_\mathrm{base}=\unit[170]{K}$.\label{fig:themGrad}}
\end{figure}

We measured the ANE by applying a current through the lithographically defined platinum heater stripe to generate a thermal gradient in the substrate via dissipation of the Joule heating, as schematized in Fig. \ref{fig:themGrad}(a). This scheme enabled us to heat one end of the sample with respect to the other end, resulting in the thermal gradient. We then used the lithographically defined thermometers to measure the temperature drop between the "cold" and "hot" ends of the device or, more generally, as a function of position on the surface  (the spatial variation in the temperature gradient will be discussed later). We define the temperature gradient as $\nabla_x T=\frac{\Delta T}{d}$, where $d$ is the distance between the two thermometers. We also measured the temperature as a function of position on the sample using two transversal arms of the Hall bar as resistive temperature sensors (see Fig. \ref{fig:themGrad}(a)), which resulted in a  more local evaluation of the temperature gradient because we evaluated the temperature at the Hall bar and not at the position of the thermometers. We observed good agreement between the results of the two thermometry methods. The thermal gradient was experimentally measured at various base (cryostat) temperatures using various heater powers (Fig. \ref{fig:themGrad}(b)). Note that we also performed the experiments using two different cryostats with various chip carriers, sample mountings and helium pressures in the variable temperature insert (VTI) of the cryostats. Comparing the results obtained using the two setups, as well as the two thermometry methods, showed that the different experimental conditions result in differences in the thermal gradient on the order of only \unit[20]{\%}. Therefore, the two experimental setups could be modeled by one set of parameters in our simulations. We present the results obtained using the various measurement setups and the corresponding thermometry below.
The spatial variation in the thermal gradient depends strongly on the sample material parameters. We modeled the heat transfer in the device including the substrate, glue, and chip carrier, using a realistic geometry and material-specific thermal conductivities. We employed the finite element method to solve the model and determine the spatial dependence of the thermal gradient. The model includes the charge current through the heater, which provides Joule heating as an input to the thermal component. The conductivity of the thin platinum wire was obtained by fitting to the measured heater power of \unit[0.26]{W} at fixed base temperature of \unit[170]{K}. The conductivity of \unitfrac[1040500] {S}{m} is lower than for bulk Pt as expected. We used a thermal conductivity of \unitfrac[12]{W}{Km}  for the STO substrate \cite{dehkordi2018}. As the STO substrate is glued onto a ceramic chip carrier, which is clamped to a copper heat sink, a \unit[0.2]{mm} thick layer with a thermal conductivity of \unitfrac[0.3]{W}{Km} was added between the substrate and the chip carrier as thermal resistance of the glue in our simulations. Using the thermal conductivity of the \unit[0.635]{mm} thick chip carrier as a fitting parameter, yielded to a value of \unitfrac[0.37]{W}{Km}. The temperature of the hottest transversal bar defined the fitting criterion. The simulation results show that the in-plane thermal gradient depends strongly on the thermal conductivity of the substrate, as well as on the thermal contact of the substrate with the heat sink. The values of the thermal conductivities and heat capacities used in the simulation are given in Tab. \ref{tab:simu_param}. Fig. \ref{fig:themGrad}(c) shows the simulated temperature profile for the sample surface. In Fig. \ref{fig:themGrad}(d) we show the resulting temperature profile and gradient along the $x$ direction at $y=\unit[0]{mm}$ as blue and green curves, respectively. The resulting simulated difference of the temperatures averaged along the hottest and coldest transversal bar at a base temperature of \unit[170]{K} and a heater power of \unit[0.26]{W} is \unit[5.88]{K}, whereas the measured global temperature difference is \unit[4.16]{K} under the same conditions. As we only use the shape of the decay of the temperature gradient along $y$ to rescale our measured temperature difference, this level of agreement is sufficient. 

\begin{table}[H]
    \centering
    \begin{tabular}{|c|c|c|}
    \hline
        Material & \makecell{Thermal conductivity\\(\unitfrac{W}{Km})}  & \makecell{Heat capacity\\(\unitfrac{J}{K\,kg})} \\ \hline
        Mn$_3$NiN (film) & 17 & 600  \\
        SrTiO$_3$ (substrate) & 12 & 500 \\
        GE varnish (glue) & 0.3 & 500 \\
        Al$_2$O$_3$ (chip carrier, fit parameter) & 0.37 & 730 \\
        Pt (heater and thermometers) & 71.6 & 133 \\\hline
\end{tabular}
    \caption{Overview of the parameters used in the FEM simulations.}
    \label{tab:simu_param}
\end{table}

\begin{figure}[H]
	\centering
	\includegraphics[width=\textwidth]{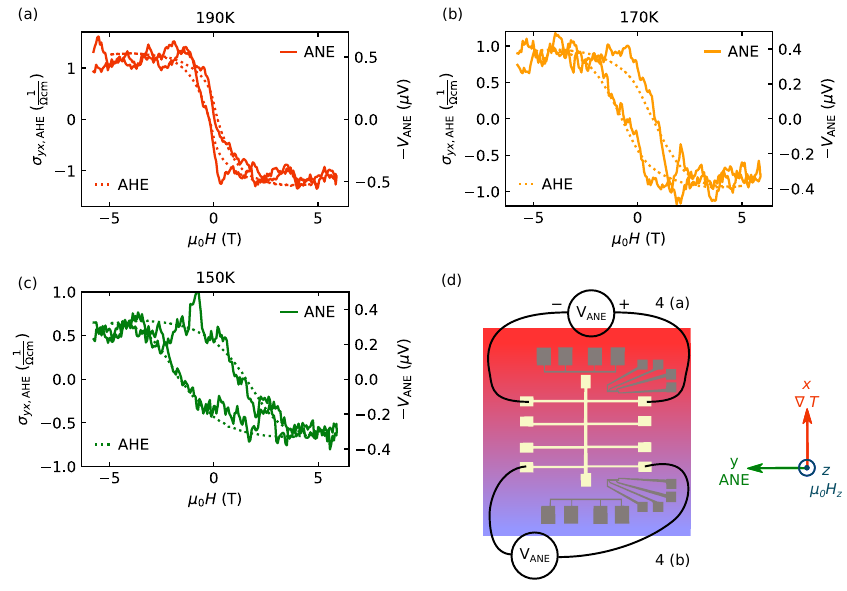}
	\caption{\textbf{Comparison of the field dependence of the AHE (dotted curves) and ANE (solid curve) curves.} Base temperatures (a) \unit[190]{K}; (b) \unit[170]{K} and (c) \unit[150]{K}. (d) Schematic of the sample setup used to measure the anomalous Nernst effect. Upper heater at $x=\unit[0.963]{mm}$, upper transversal bar at $x=\unit[0.5]{mm}$ and lower transversal bar at $x=\unit[-0.5]{mm}$. \label{fig:AHE_ANE}}
\end{figure}

We determined the anomalous Nernst voltage at various pairs of transverse contacts to experimentally explore the spatial dependence of the applied temperature gradient $\nabla_x T$. The magnetic field was applied out of the sample plane (Fig. \ref{fig:AHE_ANE}(d)). The ANE signal was measured at base temperatures of $T_\mathrm{base}$ 150~K, 170~K and 190~K. Subtracting the linear background corresponding to the ordinary Nernst contribution reveals a saturation of the Nernst signal.  Fig. \ref{fig:AHE_ANE}(a) to (c) shows the measured anomalous Nernst voltages at different base temperatures $T_\mathrm{base}$ in comparison to the  anomalous Hall conductivities discussed above. The coercive field of the ANE for each base temperature $T_\mathrm{base}$ is in agreement with the values measured based on the AHE. Note that the anomalous Nernst voltage plotted in Fig. \ref{fig:AHE_ANE} depends on the applied thermal gradient, therefore, one cannot directly compare the magnitude of the ANE. Thus, to perform such a comparison, we calculate the anomalous Nernst coefficient using the local temperature gradients at our transversal bars. This procedure is presented below.

\begin{figure}[H]
	\centering
	\includegraphics[width=\textwidth]{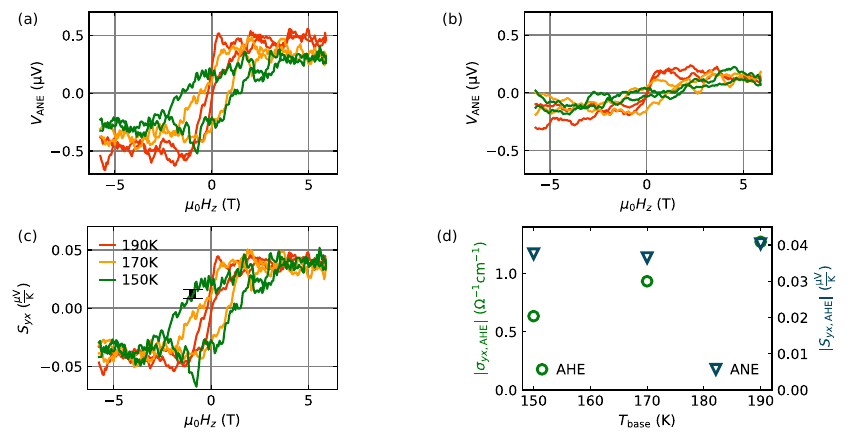}
	\caption{\textbf{Mesurements of the anomalous Nernst effect.}(a) Magnetic-field dependence of the anomalous Nernst voltage on the upper transversal bar ($x=\unit[0.5]{mm}$) for different temperatures. (b) Magnetic-field dependence of the anomalous Nernst voltage on the lower transversal bar ($x=\unit[-0.5]{mm}$) for different temperatures.	(c) Magnetic-field dependence of the ANE on the upper transversal bar for different temperatures, normalized to the mean ANE for each temperature. (d) Comparison between the temperature dependence of the AHE and the ANE.
	\label{fig:Nernst}}
\end{figure}

To properly evaluate the ANE coefficient, it is essential to assess the spatial homogeneity of the signals. For this purpose, we recorded the ANE signal for different pairs of contacts, i.e., at different distances $x$ from the heater. The data compiled in Fig. \ref{fig:Nernst}(a) and (b) show the field depended anomalous Nernst voltage for the transversal contact pair closest and furthest from the heater, respectively. The ANE voltages measured on the two contact pairs differs by a factor of approximately 2.8, confirming the inhomogeneity of the temperature gradient expected from our FEM simulation, which predicts a factor of 2.5 between the local temperature gradients at these two contact pairs. We determined the anomalous Nernst coefficient $S_{yx}$ using the following formula: 
\begin{equation}
    S_{yx}=\frac{V_y}{\nabla_x T_\mathrm{loc,meas}\times w_\mathrm{bond}}
\end{equation}
where $V_y$ is the transversal voltage between the contracts, $w_\mathrm{bond}$ is the distance between the contacts and $\nabla_x T_\mathrm{loc,meas}$ is the thermal gradient along $x$, which was corrected for the spatial variation by using the FEM simulation results. The correction was calculated in the following way. We used the simulated local temperature gradients $\nabla_x T_\mathrm{loc,simu} (x)$ and global temperature difference $\Delta T_\mathrm{glob, simu}$, as well as the measured global temperature difference $\Delta T_\mathrm{glob, meas}$, to calculate the local temperature gradients $\nabla_x T_\mathrm{loc, meas}$ for each base temperature, heater power and distance from the heater as
\begin{equation}
    \nabla_x T_\mathrm{loc, meas} = \nabla_x T_\mathrm{loc, simu}\times\frac{\Delta T_\mathrm{glob, meas}}{\Delta T_\mathrm{glob, simu}}
    \label{eq:rescaling_factor}
\end{equation}
The inhomogeneity of the temperature along the $y$-direction can also be estimated from the FEM simulation results. We considered this inhomogeneity in calculating the simulated temperature difference $\Delta T_\mathrm{glob, simu}$ and temperature gradient $\nabla_x T_\mathrm{loc,simu}$ by averaging the temperature and temperature gradient for each transversal bar over the length of the bar ( the $y$ direction).

The resulting anomalous Nernst coefficients are shown in Fig. \ref{fig:Nernst}(c). In Fig. \ref{fig:Nernst}(d), the temperature dependency of the ANE (blue) is compared to that of the AHE (green). The ANE is nearly constant within this temperature range, whereas the AHE decreases with decreasing temperature.

In addition to determine the ANE, we evaluated the Seebeck coefficient $ S_{xx}$ by measuring the voltage $V_x$ of the longitudinal bar at the investigated temperatures and the applied temperature gradient. The Seebeck coefficient is defined as \cite{park2020}
\begin{equation}
    S_{xx}=-\frac{V_x}{\Delta T}
\end{equation}

As the temperature profile of the sample was also inhomogenous, we used the FEM simulation results and the measured global temperature difference to calculate the local temperature difference $\Delta T$ between the two contacts of the longitudinal bar.

Having determined the anomalous Hall conductivity $\sigma_{yx\mathrm{,AHE}}$, the longitudinal conductivity $\sigma_{xx}$, the Seebeck coefficient $S_{xx}$, and the anomalous Nernst coefficient $S_{yx\mathrm{,ANE}}$, we finally calculated the anomalous Nernst conductivity $\alpha_{yx\mathrm{,ANE}}$ as \cite{pu2008}
\begin{equation}
    \alpha_{yx\mathrm{,ANE}}=\sigma_{xx}S_{yx\mathrm{,ANE}}+\sigma_{yx\mathrm{,AHE}}S_{xx}
\end{equation}
Note that the longitudinal conductivity $\sigma_{xx}$, the Seebeck coefficient $S_{xx}$, and the Nernst coefficient $S_{yx\mathrm{,ANE}}$ are positive, whereas the anomalous Hall conductivity $\sigma_{yx\mathrm{,AHE}}$ is negative. The anomalous Nernst conductivity $\alpha_{yx\mathrm{,ANE}}$ is then positive because $|\sigma_{xx}S_{yx}|>|\sigma_{yx}S_{xx}|$. The transport coefficients used to calculate the anomalous Nernst conductivity $\alpha_{yx}$ are shown in Table \ref{tab:coeff}.

\begin{table}[H]
    \centering
    \begin{tabular}{|c|c|c|c|c|c|}
    \hline
        $T$ (K) & $\sigma_{xx}$ ($\Omega^{-1}\mathrm{cm}^{-1}$) & $\sigma_{yx\mathrm{,AHE}}$ ($\Omega^{-1}\mathrm{cm}^{-1}$) & $S_{xx}$ (\unitfrac{$\mu$V}{K}) & $S_{yx\mathrm{,ANE}}$ (\unitfrac{$\mu$V}{K})& $\alpha_{yx\mathrm{,ANE}}$ (\unitfrac{A}{Km})\\ \hline
        \unit[190]{K} & $974.93$ & $-1.272$ & $1.70$ & $0.0404(23)$ & $0.00372(22)$\\
        \unit[170]{K} & $964.76$ & $-0.933$  & $2.39$  &$0.0365(23)$ & $0.00330(22)$ \\
        \unit[150]{K} & $956.91$ & $-0.634$ & $2.94$  & $0.0376(39)$ & $0.00341(37)$ \\ \hline
\end{tabular}
    \caption{Transport coefficients used to calculate the Nernst conductivity $\alpha_{yx\mathrm{,ANE}}$. }
    \label{tab:coeff}
\end{table}

The measured ANE conductivity of $\alpha_{yx}=\unitfrac[0.00348]{A}{Km}$ (the average of the measurements in the temperature range from \unit[150]{K} to \unit[190]{K}) is significantly smaller than the theoretically predicted $|\alpha_{yx}|=\unitfrac[1.80]{A}{Km}$ \cite{zhou2020a}. There are multiple possible reasons for this difference that are discussed below. Note that only the projection of the Hall vector onto the out-of-plane (e.g., [001]) direction was experimentally measured, whereas the theoretical value corresponds to the [111] direction. However, projecting the [111] direction on the [001] direction in a cubic crystal would only lead to a factor of $\frac{1}{\sqrt{3}}$, which would not explain the observed discrepancy of two orders of magnitude.

\begin{figure}[H]
	\centering
	\includegraphics[width=\textwidth]{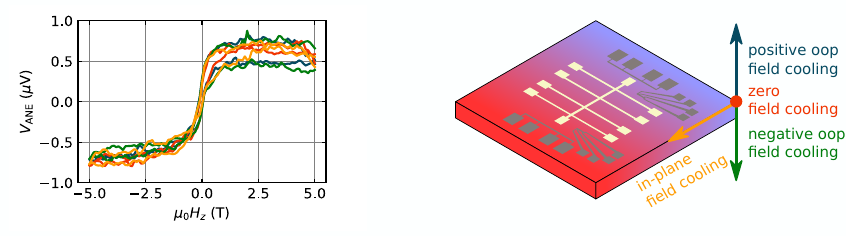}
	\caption{Anomalous Nernst voltages at $T_\mathrm{base}=\unit[190]{K}$ for different field-cooling orientations and without field cooling. Field cooling was performed by cooling the sample from \unit[300]{K} to \unit[190]{K} in a field of \unit[5]{T} with different orientations (blue, yellow and green curve) and with no field (red curve).\label{fig:Fieldcooled}}
\end{figure}

Another possible reason for the discrepancy between the experimental and theoretical results could be the multidomain character of the sample. Both $\Gamma^{4g}$ and $\Gamma^{5g}$ phases can coexist \cite{zhou2020a} in Mn$_3$NiN. The $\Gamma^{4g}$ phase alone can break into eight equivalent domains. The Hall vectors and net moments corresponding to the 8 domains have been summarized by Johnson et al. \cite{johnson2022}. The overall Nernst effect could thus be reduced due to the multidomain character of the sample. However, the domain population in an antiferromagnet can be controlled by cooling the material under a magnetic field from above the Néel temperature \cite{reichlova2019,johnson2022}. To investigate the effect of the domain structure on our ANE signal, we cooled the sample in various magnetic field orientations and in zero-field from above the Néel temperature (e. g. \unit[300]{K}) down to \unit[190]{K}. We did not observe any variation of the measured ANE signal, as demonstrated by Fig. \ref{fig:Fieldcooled}. This result suggests that the multidomain character of the sample cannot account for the small ANE coefficient.
Another possible reason for the discrepancy between the theoretical and experimental results is the strong sensitivity of the anomalous Nernst effect to the position of the Fermi level. The ANE is dominated by the Berry curvature at the Fermi level, unlike the AHE, which has contributions from all the occupied energy bands. As shown by Zhou et al. \cite{zhou2020a} in Fig. 3c  the anomalous Nernst conductivity can be reduced, eliminated or even change sign if the Fermi level is shifted by only 30~meV. The position of the Fermi level is experimentally very sensitive to the stoichiometry and purity of a particular sample. The calculations also do not include scattering, which can play a significant role in a real sample and to which the ANE may be particularly sensitive. 

Similarly, the strain induced by the substrate can strongly influence the position of the Fermi level, consequently, the ANE becomes a sensitive probe of the strain-induced changes, as You et al. \cite{you2022} have illustrated. The calculation results presented in Fig. 2(b) of a study by Boldrin et al. \cite{boldrin2019} show that strain can influence the energy dependence of the anomalous Hall conductivity and, therefore, also the ANE.

Clearly, there are several possible explanations for the discrepancy between the experimental and theoretical values of the anomalous Nernst conductivity, warranting further study. Nevertheless, Mn$_3$NiN offers an ideal platform to explore the spin-caloritronic phenomena because the ANE can be measured both by the in- and out-of-plane thermal gradients in one device simultaneously \cite{schreier2013}. This is because the non-collinear magnetic structure causes that all the off-diagonal elements of the Hall conductivity are nonzero, i.e., as the Hall vector is close to the body diagonal, both thermal gradients can generate transverse voltage. Moreover the out-of-plane thermal gradient can be applied either locally \cite{johnson2022} or globally \cite{you2022}. 

\section{Summary and Outlook}
In conclusion, we measured the anomalous Hall and Nernst effect in Mn$_3$NiN in the same sample. We used our experimental data and finite-element simulation results to obtain the anomalous Nernst coefficient in the Mn$_3$NiN thin film as $\unitfrac[0.0382] {\mu V}{K}$. The anomalous Nernst conductivity was measured as $\alpha_{yx}=\unitfrac[0.00348]{A}{Km}$ (by averaging the respective values over the temperature range from \unit[150]{K} to \unit[190]{K}), which is significantly smaller than the theoretical prediction of $|\alpha_{yx}|=\unitfrac[1.8]{A}{Km}$. To accurately determine of the ANE coefficient, we have paid special attention to characterize the spatial distribution of the thermal gradient in our sample, and we supported our measurements by FEM simulations. A likely reason for the discrepancy between the experimental and theoretical ANE values is the particular position of the Fermi level in our samples. We propose Mn$_3$NiN films with a [001] film normal as a fundamentally interesting model system to explore the rich physics of non-collinear antiferromagnets. In addition, a Mn$_3$NiN film a good model system for simultaneous in-plane and out-of-plane thermal gradient measurements to explore the anisotropy of the ANE in this class of materials.

\section{Acknowledgments}
S.B. was supported by the DFG through Project C08 of SFB 1143. This study was supported in part by the Ministry of Education of the Czech Republic Grants LM2018110 and LNSM-LNSpin, by the Grant Agency of the Czech Republic (Grant No. 22-17899K), and EU FET Open RIA (Grant No. 766566). D.K. acknowledges a Lumina Quaeruntur fellowship from the Czech Academy of Sciences. J.Z. was supported by the Ministry of Education, Youth and Sports of the Czech Republic from the OP RDE program under the project International Mobility of Researchers MSCAIF at CTU (No. CZ.02.2.69/0.0/0.0/18 070/0010457) and through the e-INFRA CZ (ID:90140). E.S. and P.N. would like to acknowledge a Czech Science Foundation (Grant no. 19-28375X).

\section*{References}


\begin{thebibliography}{33}%
\makeatletter
\providecommand \@ifxundefined [1]{%
 \@ifx{#1\undefined}
}%
\providecommand \@ifnum [1]{%
 \ifnum #1\expandafter \@firstoftwo
 \else \expandafter \@secondoftwo
 \fi
}%
\providecommand \@ifx [1]{%
 \ifx #1\expandafter \@firstoftwo
 \else \expandafter \@secondoftwo
 \fi
}%
\providecommand \natexlab [1]{#1}%
\providecommand \enquote  [1]{``#1''}%
\providecommand \bibnamefont  [1]{#1}%
\providecommand \bibfnamefont [1]{#1}%
\providecommand \citenamefont [1]{#1}%
\providecommand \href@noop [0]{\@secondoftwo}%
\providecommand \href [0]{\begingroup \@sanitize@url \@href}%
\providecommand \@href[1]{\@@startlink{#1}\@@href}%
\providecommand \@@href[1]{\endgroup#1\@@endlink}%
\providecommand \@sanitize@url [0]{\catcode `\\12\catcode `\$12\catcode
  `\&12\catcode `\#12\catcode `\^12\catcode `\_12\catcode `\%12\relax}%
\providecommand \@@startlink[1]{}%
\providecommand \@@endlink[0]{}%
\providecommand \url  [0]{\begingroup\@sanitize@url \@url }%
\providecommand \@url [1]{\endgroup\@href {#1}{\urlprefix }}%
\providecommand \urlprefix  [0]{URL }%
\providecommand \Eprint [0]{\href }%
\providecommand \doibase [0]{https://doi.org/}%
\providecommand \selectlanguage [0]{\@gobble}%
\providecommand \bibinfo  [0]{\@secondoftwo}%
\providecommand \bibfield  [0]{\@secondoftwo}%
\providecommand \translation [1]{[#1]}%
\providecommand \BibitemOpen [0]{}%
\providecommand \bibitemStop [0]{}%
\providecommand \bibitemNoStop [0]{.\EOS\space}%
\providecommand \EOS [0]{\spacefactor3000\relax}%
\providecommand \BibitemShut  [1]{\csname bibitem#1\endcsname}%
\let\auto@bib@innerbib\@empty
\bibitem [{\citenamefont {Ding}\ \emph {et~al.}(2019)\citenamefont {Ding},
  \citenamefont {Koo}, \citenamefont {Xu}, \citenamefont {Li}, \citenamefont
  {Lu}, \citenamefont {Zhao}, \citenamefont {Wang}, \citenamefont {Yin},
  \citenamefont {Lei}, \citenamefont {Yan}, \citenamefont {Zhu},\ and\
  \citenamefont {Behnia}}]{ding2019}%
  \BibitemOpen
  \bibfield  {author} {\bibinfo {author} {\bibfnamefont {L.}~\bibnamefont
  {Ding}}, \bibinfo {author} {\bibfnamefont {J.}~\bibnamefont {Koo}}, \bibinfo
  {author} {\bibfnamefont {L.}~\bibnamefont {Xu}}, \bibinfo {author}
  {\bibfnamefont {X.}~\bibnamefont {Li}}, \bibinfo {author} {\bibfnamefont
  {X.}~\bibnamefont {Lu}}, \bibinfo {author} {\bibfnamefont {L.}~\bibnamefont
  {Zhao}}, \bibinfo {author} {\bibfnamefont {Q.}~\bibnamefont {Wang}}, \bibinfo
  {author} {\bibfnamefont {Q.}~\bibnamefont {Yin}}, \bibinfo {author}
  {\bibfnamefont {H.}~\bibnamefont {Lei}}, \bibinfo {author} {\bibfnamefont
  {B.}~\bibnamefont {Yan}}, \bibinfo {author} {\bibfnamefont {Z.}~\bibnamefont
  {Zhu}},\ and\ \bibinfo {author} {\bibfnamefont {K.}~\bibnamefont {Behnia}},\
  }\bibfield  {title} {\bibinfo {title} {{Intrinsic Anomalous Nernst Effect
  Amplified by Disorder in a Half-Metallic Semimetal}},\ }\href
  {https://doi.org/10.1103/PhysRevX.9.041061} {\bibfield  {journal} {\bibinfo
  {journal} {Phys. Rev. X}\ }\textbf {\bibinfo {volume} {9}},\ \bibinfo {pages}
  {041061} (\bibinfo {year} {2019})}\BibitemShut {NoStop}%
\bibitem [{\citenamefont {Chen}\ \emph {et~al.}(2014)\citenamefont {Chen},
  \citenamefont {Niu},\ and\ \citenamefont {MacDonald}}]{chen2014}%
  \BibitemOpen
  \bibfield  {author} {\bibinfo {author} {\bibfnamefont {H.}~\bibnamefont
  {Chen}}, \bibinfo {author} {\bibfnamefont {Q.}~\bibnamefont {Niu}},\ and\
  \bibinfo {author} {\bibfnamefont {A.~H.}\ \bibnamefont {MacDonald}},\
  }\bibfield  {title} {\bibinfo {title} {{Anomalous Hall Effect Arising from
  Noncollinear Antiferromagnetism}},\ }\href
  {https://doi.org/10.1103/PhysRevLett.112.017205} {\bibfield  {journal}
  {\bibinfo  {journal} {Phys. Rev. Lett.}\ }\textbf {\bibinfo {volume} {112}},\
  \bibinfo {pages} {017205} (\bibinfo {year} {2014})}\BibitemShut {NoStop}%
\bibitem [{\citenamefont {Ikhlas}\ \emph {et~al.}(2017)\citenamefont {Ikhlas},
  \citenamefont {Tomita}, \citenamefont {Koretsune}, \citenamefont {Suzuki},
  \citenamefont {Nishio-Hamane}, \citenamefont {Arita}, \citenamefont {Otani},\
  and\ \citenamefont {Nakatsuji}}]{ikhlas2017}%
  \BibitemOpen
  \bibfield  {author} {\bibinfo {author} {\bibfnamefont {M.}~\bibnamefont
  {Ikhlas}}, \bibinfo {author} {\bibfnamefont {T.}~\bibnamefont {Tomita}},
  \bibinfo {author} {\bibfnamefont {T.}~\bibnamefont {Koretsune}}, \bibinfo
  {author} {\bibfnamefont {M.-T.}\ \bibnamefont {Suzuki}}, \bibinfo {author}
  {\bibfnamefont {D.}~\bibnamefont {Nishio-Hamane}}, \bibinfo {author}
  {\bibfnamefont {R.}~\bibnamefont {Arita}}, \bibinfo {author} {\bibfnamefont
  {Y.}~\bibnamefont {Otani}},\ and\ \bibinfo {author} {\bibfnamefont
  {S.}~\bibnamefont {Nakatsuji}},\ }\bibfield  {title} {\bibinfo {title}
  {{Large anomalous Nernst effect at room temperature in a chiral
  antiferromagnet}},\ }\href {https://doi.org/10.1038/nphys4181} {\bibfield
  {journal} {\bibinfo  {journal} {Nat. Phys.}\ }\textbf {\bibinfo {volume}
  {13}},\ \bibinfo {pages} {1085} (\bibinfo {year} {2017})}\BibitemShut
  {NoStop}%
\bibitem [{\citenamefont {Chen}\ \emph {et~al.}(2022)\citenamefont {Chen},
  \citenamefont {Minami}, \citenamefont {Sakai}, \citenamefont {Wang},
  \citenamefont {Feng}, \citenamefont {Nomoto}, \citenamefont {Hirayama},
  \citenamefont {Ishii}, \citenamefont {Koretsune}, \citenamefont {Arita},\
  and\ \citenamefont {Nakatsuji}}]{chen2022}%
  \BibitemOpen
  \bibfield  {author} {\bibinfo {author} {\bibfnamefont {T.}~\bibnamefont
  {Chen}}, \bibinfo {author} {\bibfnamefont {S.}~\bibnamefont {Minami}},
  \bibinfo {author} {\bibfnamefont {A.}~\bibnamefont {Sakai}}, \bibinfo
  {author} {\bibfnamefont {Y.}~\bibnamefont {Wang}}, \bibinfo {author}
  {\bibfnamefont {Z.}~\bibnamefont {Feng}}, \bibinfo {author} {\bibfnamefont
  {T.}~\bibnamefont {Nomoto}}, \bibinfo {author} {\bibfnamefont
  {M.}~\bibnamefont {Hirayama}}, \bibinfo {author} {\bibfnamefont
  {R.}~\bibnamefont {Ishii}}, \bibinfo {author} {\bibfnamefont
  {T.}~\bibnamefont {Koretsune}}, \bibinfo {author} {\bibfnamefont
  {R.}~\bibnamefont {Arita}},\ and\ \bibinfo {author} {\bibfnamefont
  {S.}~\bibnamefont {Nakatsuji}},\ }\bibfield  {title} {\bibinfo {title}
  {{Large anomalous Nernst effect and nodal plane in an iron-based kagome
  ferromagnet}},\ }\href {https://doi.org/10.1126/sciadv.abk1480} {\bibfield
  {journal} {\bibinfo  {journal} {Sci. Adv.}\ }\textbf {\bibinfo {volume}
  {8}},\ \bibinfo {pages} {eabk1480} (\bibinfo {year} {2022})}\BibitemShut
  {NoStop}%
\bibitem [{\citenamefont {Asaba}\ \emph {et~al.}(2021)\citenamefont {Asaba},
  \citenamefont {Ivanov}, \citenamefont {Thomas}, \citenamefont {Savrasov},
  \citenamefont {Thompson}, \citenamefont {Bauer},\ and\ \citenamefont
  {Ronning}}]{asaba2021}%
  \BibitemOpen
  \bibfield  {author} {\bibinfo {author} {\bibfnamefont {T.}~\bibnamefont
  {Asaba}}, \bibinfo {author} {\bibfnamefont {V.}~\bibnamefont {Ivanov}},
  \bibinfo {author} {\bibfnamefont {S.~M.}\ \bibnamefont {Thomas}}, \bibinfo
  {author} {\bibfnamefont {S.~Y.}\ \bibnamefont {Savrasov}}, \bibinfo {author}
  {\bibfnamefont {J.~D.}\ \bibnamefont {Thompson}}, \bibinfo {author}
  {\bibfnamefont {E.~D.}\ \bibnamefont {Bauer}},\ and\ \bibinfo {author}
  {\bibfnamefont {F.}~\bibnamefont {Ronning}},\ }\bibfield  {title} {\bibinfo
  {title} {{Colossal anomalous Nernst effect in a correlated noncentrosymmetric
  kagome ferromagnet}},\ }\href {https://doi.org/10.1126/sciadv.abf1467}
  {\bibfield  {journal} {\bibinfo  {journal} {Sci. Adv.}\ }\textbf {\bibinfo
  {volume} {7}},\ \bibinfo {pages} {eabf1467} (\bibinfo {year}
  {2021})}\BibitemShut {NoStop}%
\bibitem [{\citenamefont {Pan}\ \emph {et~al.}(2022)\citenamefont {Pan},
  \citenamefont {Le}, \citenamefont {He}, \citenamefont {Watzman},
  \citenamefont {Yao}, \citenamefont {Gooth}, \citenamefont {Heremans},
  \citenamefont {Sun},\ and\ \citenamefont {Felser}}]{pan2022}%
  \BibitemOpen
  \bibfield  {author} {\bibinfo {author} {\bibfnamefont {Y.}~\bibnamefont
  {Pan}}, \bibinfo {author} {\bibfnamefont {C.}~\bibnamefont {Le}}, \bibinfo
  {author} {\bibfnamefont {B.}~\bibnamefont {He}}, \bibinfo {author}
  {\bibfnamefont {S.~J.}\ \bibnamefont {Watzman}}, \bibinfo {author}
  {\bibfnamefont {M.}~\bibnamefont {Yao}}, \bibinfo {author} {\bibfnamefont
  {J.}~\bibnamefont {Gooth}}, \bibinfo {author} {\bibfnamefont {J.~P.}\
  \bibnamefont {Heremans}}, \bibinfo {author} {\bibfnamefont {Y.}~\bibnamefont
  {Sun}},\ and\ \bibinfo {author} {\bibfnamefont {C.}~\bibnamefont {Felser}},\
  }\bibfield  {title} {\bibinfo {title} {{Giant anomalous Nernst signal in the
  antiferromagnet YbMnBi$_2$}},\ }\href
  {https://doi.org/10.1038/s41563-021-01149-2} {\bibfield  {journal} {\bibinfo
  {journal} {Nat. Mater.}\ }\textbf {\bibinfo {volume} {21}},\ \bibinfo {pages}
  {203} (\bibinfo {year} {2022})}\BibitemShut {NoStop}%
\bibitem [{\citenamefont {Zhou}\ \emph {et~al.}(2020)\citenamefont {Zhou},
  \citenamefont {Hanke}, \citenamefont {Feng}, \citenamefont {Blügel},
  \citenamefont {Mokrousov},\ and\ \citenamefont {Yao}}]{zhou2020a}%
  \BibitemOpen
  \bibfield  {author} {\bibinfo {author} {\bibfnamefont {X.}~\bibnamefont
  {Zhou}}, \bibinfo {author} {\bibfnamefont {J.-P.}\ \bibnamefont {Hanke}},
  \bibinfo {author} {\bibfnamefont {W.}~\bibnamefont {Feng}}, \bibinfo {author}
  {\bibfnamefont {S.}~\bibnamefont {Blügel}}, \bibinfo {author} {\bibfnamefont
  {Y.}~\bibnamefont {Mokrousov}},\ and\ \bibinfo {author} {\bibfnamefont
  {Y.}~\bibnamefont {Yao}},\ }\bibfield  {title} {\bibinfo {title} {{Giant
  anomalous Nernst effect in noncollinear antiferromagnetic Mn-based
  antiperovskite nitrides}},\ }\href
  {https://doi.org/10.1103/PhysRevMaterials.4.024408} {\bibfield  {journal}
  {\bibinfo  {journal} {Phys. Rev. Materials}\ }\textbf {\bibinfo {volume}
  {4}},\ \bibinfo {pages} {024408} (\bibinfo {year} {2020})}\BibitemShut
  {NoStop}%
\bibitem [{\citenamefont {Wuttke}\ \emph {et~al.}(2019)\citenamefont {Wuttke},
  \citenamefont {Caglieris}, \citenamefont {Sykora}, \citenamefont
  {Scaravaggi}, \citenamefont {Wolter}, \citenamefont {Manna}, \citenamefont
  {Süss}, \citenamefont {Shekhar}, \citenamefont {Felser}, \citenamefont
  {Büchner},\ and\ \citenamefont {Hess}}]{wuttke2019}%
  \BibitemOpen
  \bibfield  {author} {\bibinfo {author} {\bibfnamefont {C.}~\bibnamefont
  {Wuttke}}, \bibinfo {author} {\bibfnamefont {F.}~\bibnamefont {Caglieris}},
  \bibinfo {author} {\bibfnamefont {S.}~\bibnamefont {Sykora}}, \bibinfo
  {author} {\bibfnamefont {F.}~\bibnamefont {Scaravaggi}}, \bibinfo {author}
  {\bibfnamefont {A.~U.~B.}\ \bibnamefont {Wolter}}, \bibinfo {author}
  {\bibfnamefont {K.}~\bibnamefont {Manna}}, \bibinfo {author} {\bibfnamefont
  {V.}~\bibnamefont {Süss}}, \bibinfo {author} {\bibfnamefont
  {C.}~\bibnamefont {Shekhar}}, \bibinfo {author} {\bibfnamefont
  {C.}~\bibnamefont {Felser}}, \bibinfo {author} {\bibfnamefont
  {B.}~\bibnamefont {Büchner}},\ and\ \bibinfo {author} {\bibfnamefont
  {C.}~\bibnamefont {Hess}},\ }\bibfield  {title} {\bibinfo {title} {{Berry
  curvature unravelled by the anomalous Nernst effect in Mn$_3$Ge}},\ }\href
  {https://doi.org/10.1103/PhysRevB.100.085111} {\bibfield  {journal} {\bibinfo
   {journal} {Phys. Rev. B}\ }\textbf {\bibinfo {volume} {100}},\ \bibinfo
  {pages} {085111} (\bibinfo {year} {2019})}\BibitemShut {NoStop}%
\bibitem [{\citenamefont {Noky}\ \emph {et~al.}(2018)\citenamefont {Noky},
  \citenamefont {Gooth}, \citenamefont {Felser},\ and\ \citenamefont
  {Sun}}]{noky2018}%
  \BibitemOpen
  \bibfield  {author} {\bibinfo {author} {\bibfnamefont {J.}~\bibnamefont
  {Noky}}, \bibinfo {author} {\bibfnamefont {J.}~\bibnamefont {Gooth}},
  \bibinfo {author} {\bibfnamefont {C.}~\bibnamefont {Felser}},\ and\ \bibinfo
  {author} {\bibfnamefont {Y.}~\bibnamefont {Sun}},\ }\bibfield  {title}
  {\bibinfo {title} {{Characterization of topological band structures away from
  the Fermi level by the anomalous Nernst effect}},\ }\href
  {https://doi.org/10.1103/PhysRevB.98.241106} {\bibfield  {journal} {\bibinfo
  {journal} {Phys. Rev. B}\ }\textbf {\bibinfo {volume} {98}},\ \bibinfo
  {pages} {241106} (\bibinfo {year} {2018})}\BibitemShut {NoStop}%
\bibitem [{\citenamefont {Sun}\ \emph {et~al.}(1996)\citenamefont {Sun},
  \citenamefont {Gallagher}, \citenamefont {Duncombe}, \citenamefont
  {Krusin‐Elbaum}, \citenamefont {Altman}, \citenamefont {Gupta},
  \citenamefont {Lu}, \citenamefont {Gong},\ and\ \citenamefont
  {Xiao}}]{sun1996}%
  \BibitemOpen
  \bibfield  {author} {\bibinfo {author} {\bibfnamefont {J.~Z.}\ \bibnamefont
  {Sun}}, \bibinfo {author} {\bibfnamefont {W.~J.}\ \bibnamefont {Gallagher}},
  \bibinfo {author} {\bibfnamefont {P.~R.}\ \bibnamefont {Duncombe}}, \bibinfo
  {author} {\bibfnamefont {L.}~\bibnamefont {Krusin‐Elbaum}}, \bibinfo
  {author} {\bibfnamefont {R.~A.}\ \bibnamefont {Altman}}, \bibinfo {author}
  {\bibfnamefont {A.}~\bibnamefont {Gupta}}, \bibinfo {author} {\bibfnamefont
  {Y.}~\bibnamefont {Lu}}, \bibinfo {author} {\bibfnamefont {G.~Q.}\
  \bibnamefont {Gong}},\ and\ \bibinfo {author} {\bibfnamefont
  {G.}~\bibnamefont {Xiao}},\ }\bibfield  {title} {\bibinfo {title}
  {{{Observation of large low‐field magnetoresistance in trilayer
  perpendicular transport devices made using doped manganate perovskites}}},\
  }\href {https://doi.org/10.1063/1.118031} {\bibfield  {journal} {\bibinfo
  {journal} {Appl. Phys. Lett.}\ }\textbf {\bibinfo {volume} {69}},\ \bibinfo
  {pages} {3266} (\bibinfo {year} {1996})}\BibitemShut {NoStop}%
\bibitem [{\citenamefont {Reichlova}\ \emph {et~al.}(2019)\citenamefont
  {Reichlova}, \citenamefont {Janda}, \citenamefont {Godinho}, \citenamefont
  {Markou}, \citenamefont {Kriegner}, \citenamefont {Schlitz}, \citenamefont
  {Zelezny}, \citenamefont {Soban}, \citenamefont {Bejarano}, \citenamefont
  {Schultheiss}, \citenamefont {Nemec}, \citenamefont {Jungwirth},
  \citenamefont {Felser}, \citenamefont {Wunderlich},\ and\ \citenamefont
  {Goennenwein}}]{reichlova2019}%
  \BibitemOpen
  \bibfield  {author} {\bibinfo {author} {\bibfnamefont {H.}~\bibnamefont
  {Reichlova}}, \bibinfo {author} {\bibfnamefont {T.}~\bibnamefont {Janda}},
  \bibinfo {author} {\bibfnamefont {J.}~\bibnamefont {Godinho}}, \bibinfo
  {author} {\bibfnamefont {A.}~\bibnamefont {Markou}}, \bibinfo {author}
  {\bibfnamefont {D.}~\bibnamefont {Kriegner}}, \bibinfo {author}
  {\bibfnamefont {R.}~\bibnamefont {Schlitz}}, \bibinfo {author} {\bibfnamefont
  {J.}~\bibnamefont {Zelezny}}, \bibinfo {author} {\bibfnamefont
  {Z.}~\bibnamefont {Soban}}, \bibinfo {author} {\bibfnamefont
  {M.}~\bibnamefont {Bejarano}}, \bibinfo {author} {\bibfnamefont
  {H.}~\bibnamefont {Schultheiss}}, \bibinfo {author} {\bibfnamefont
  {P.}~\bibnamefont {Nemec}}, \bibinfo {author} {\bibfnamefont
  {T.}~\bibnamefont {Jungwirth}}, \bibinfo {author} {\bibfnamefont
  {C.}~\bibnamefont {Felser}}, \bibinfo {author} {\bibfnamefont
  {J.}~\bibnamefont {Wunderlich}},\ and\ \bibinfo {author} {\bibfnamefont
  {S.~T.~B.}\ \bibnamefont {Goennenwein}},\ }\bibfield  {title} {\bibinfo
  {title} {{Imaging and writing magnetic domains in the non-collinear
  antiferromagnet Mn$_3$Sn}},\ }\href
  {https://doi.org/10.1038/s41467-019-13391-z} {\bibfield  {journal} {\bibinfo
  {journal} {Nat. Commun.}\ }\textbf {\bibinfo {volume} {10}},\ \bibinfo
  {pages} {5459} (\bibinfo {year} {2019})}\BibitemShut {NoStop}%
\bibitem [{\citenamefont {Janda}\ \emph {et~al.}(2020)\citenamefont {Janda},
  \citenamefont {Godinho}, \citenamefont {Ostatnicky}, \citenamefont
  {Pfitzner}, \citenamefont {Ulrich}, \citenamefont {Hoehl}, \citenamefont
  {Reimers}, \citenamefont {Šobáň}, \citenamefont {Metzger}, \citenamefont
  {Reichlová}, \citenamefont {Novák}, \citenamefont {Campion}, \citenamefont
  {Heberle}, \citenamefont {Wadley}, \citenamefont {Edmonds}, \citenamefont
  {Amin}, \citenamefont {Chauhan}, \citenamefont {Dhesi}, \citenamefont
  {Maccherozzi}, \citenamefont {Otxoa}, \citenamefont {Roy}, \citenamefont
  {Olejník}, \citenamefont {Němec}, \citenamefont {Jungwirth}, \citenamefont
  {Kaestner},\ and\ \citenamefont {Wunderlich}}]{janda2020}%
  \BibitemOpen
  \bibfield  {author} {\bibinfo {author} {\bibfnamefont {T.}~\bibnamefont
  {Janda}}, \bibinfo {author} {\bibfnamefont {J.}~\bibnamefont {Godinho}},
  \bibinfo {author} {\bibfnamefont {T.}~\bibnamefont {Ostatnicky}}, \bibinfo
  {author} {\bibfnamefont {E.}~\bibnamefont {Pfitzner}}, \bibinfo {author}
  {\bibfnamefont {G.}~\bibnamefont {Ulrich}}, \bibinfo {author} {\bibfnamefont
  {A.}~\bibnamefont {Hoehl}}, \bibinfo {author} {\bibfnamefont
  {S.}~\bibnamefont {Reimers}}, \bibinfo {author} {\bibfnamefont
  {Z.}~\bibnamefont {Šobáň}}, \bibinfo {author} {\bibfnamefont
  {T.}~\bibnamefont {Metzger}}, \bibinfo {author} {\bibfnamefont
  {H.}~\bibnamefont {Reichlová}}, \bibinfo {author} {\bibfnamefont
  {V.}~\bibnamefont {Novák}}, \bibinfo {author} {\bibfnamefont {R.~P.}\
  \bibnamefont {Campion}}, \bibinfo {author} {\bibfnamefont {J.}~\bibnamefont
  {Heberle}}, \bibinfo {author} {\bibfnamefont {P.}~\bibnamefont {Wadley}},
  \bibinfo {author} {\bibfnamefont {K.~W.}\ \bibnamefont {Edmonds}}, \bibinfo
  {author} {\bibfnamefont {O.~J.}\ \bibnamefont {Amin}}, \bibinfo {author}
  {\bibfnamefont {J.~S.}\ \bibnamefont {Chauhan}}, \bibinfo {author}
  {\bibfnamefont {S.~S.}\ \bibnamefont {Dhesi}}, \bibinfo {author}
  {\bibfnamefont {F.}~\bibnamefont {Maccherozzi}}, \bibinfo {author}
  {\bibfnamefont {R.~M.}\ \bibnamefont {Otxoa}}, \bibinfo {author}
  {\bibfnamefont {P.~E.}\ \bibnamefont {Roy}}, \bibinfo {author} {\bibfnamefont
  {K.}~\bibnamefont {Olejník}}, \bibinfo {author} {\bibfnamefont
  {P.}~\bibnamefont {Němec}}, \bibinfo {author} {\bibfnamefont
  {T.}~\bibnamefont {Jungwirth}}, \bibinfo {author} {\bibfnamefont
  {B.}~\bibnamefont {Kaestner}},\ and\ \bibinfo {author} {\bibfnamefont
  {J.}~\bibnamefont {Wunderlich}},\ }\bibfield  {title} {\bibinfo {title}
  {Magneto-seebeck microscopy of domain switching in collinear antiferromagnet
  {CuMnAs}},\ }\href {https://doi.org/10.1103/PhysRevMaterials.4.094413}
  {\bibfield  {journal} {\bibinfo  {journal} {Phys. Rev. Materials}\ }\textbf
  {\bibinfo {volume} {4}},\ \bibinfo {pages} {094413} (\bibinfo {year}
  {2020})}\BibitemShut {NoStop}%
\bibitem [{\citenamefont {Johnson}\ \emph {et~al.}(2022)\citenamefont
  {Johnson}, \citenamefont {Kimák}, \citenamefont {Zemen}, \citenamefont
  {Šobáň}, \citenamefont {Schmoranzerová}, \citenamefont {Godinho},
  \citenamefont {Němec}, \citenamefont {Beckert}, \citenamefont {Reichlová},
  \citenamefont {Boldrin}, \citenamefont {Wunderlich},\ and\ \citenamefont
  {Cohen}}]{johnson2022}%
  \BibitemOpen
  \bibfield  {author} {\bibinfo {author} {\bibfnamefont {F.}~\bibnamefont
  {Johnson}}, \bibinfo {author} {\bibfnamefont {J.}~\bibnamefont {Kimák}},
  \bibinfo {author} {\bibfnamefont {J.}~\bibnamefont {Zemen}}, \bibinfo
  {author} {\bibfnamefont {Z.}~\bibnamefont {Šobáň}}, \bibinfo {author}
  {\bibfnamefont {E.}~\bibnamefont {Schmoranzerová}}, \bibinfo {author}
  {\bibfnamefont {J.}~\bibnamefont {Godinho}}, \bibinfo {author} {\bibfnamefont
  {P.}~\bibnamefont {Němec}}, \bibinfo {author} {\bibfnamefont
  {S.}~\bibnamefont {Beckert}}, \bibinfo {author} {\bibfnamefont
  {H.}~\bibnamefont {Reichlová}}, \bibinfo {author} {\bibfnamefont
  {D.}~\bibnamefont {Boldrin}}, \bibinfo {author} {\bibfnamefont
  {J.}~\bibnamefont {Wunderlich}},\ and\ \bibinfo {author} {\bibfnamefont
  {L.~F.}\ \bibnamefont {Cohen}},\ }\bibfield  {title} {\bibinfo {title}
  {{Identifying the octupole antiferromagnetic domain orientation in Mn$_3$NiN
  by scanning anomalous Nernst effect microscopy}},\ }\href
  {https://doi.org/10.1063/5.0091257} {\bibfield  {journal} {\bibinfo
  {journal} {Applied Physics Letters}\ }\textbf {\bibinfo {volume} {120}},\
  \bibinfo {pages} {232402} (\bibinfo {year} {2022})}\BibitemShut {NoStop}%
\bibitem [{\citenamefont {Zhou}\ and\ \citenamefont
  {Sakuraba}(2020)}]{zhou2020}%
  \BibitemOpen
  \bibfield  {author} {\bibinfo {author} {\bibfnamefont {W.}~\bibnamefont
  {Zhou}}\ and\ \bibinfo {author} {\bibfnamefont {Y.}~\bibnamefont
  {Sakuraba}},\ }\bibfield  {title} {\bibinfo {title} {{Heat flux sensing by
  anomalous Nernst effect in Fe–Al thin films on a flexible substrate}},\
  }\href {https://doi.org/10.35848/1882-0786/ab79fe} {\bibfield  {journal}
  {\bibinfo  {journal} {Appl. Phys. Express}\ }\textbf {\bibinfo {volume}
  {13}},\ \bibinfo {pages} {043001} (\bibinfo {year} {2020})}\BibitemShut
  {NoStop}%
\bibitem [{\citenamefont {Nan}\ \emph {et~al.}(2020)\citenamefont {Nan},
  \citenamefont {Quintela}, \citenamefont {Irwin}, \citenamefont {Gurung},
  \citenamefont {Shao}, \citenamefont {Gibbons}, \citenamefont {Campbell},
  \citenamefont {Song}, \citenamefont {Choi}, \citenamefont {Guo},
  \citenamefont {Johnson}, \citenamefont {Manuel}, \citenamefont {Chopdekar},
  \citenamefont {Hallsteinsen}, \citenamefont {Tybell}, \citenamefont {Ryan},
  \citenamefont {Kim}, \citenamefont {Choi}, \citenamefont {Radaelli},
  \citenamefont {Ralph}, \citenamefont {Tsymbal}, \citenamefont {Rzchowski},\
  and\ \citenamefont {Eom}}]{nan2020}%
  \BibitemOpen
  \bibfield  {author} {\bibinfo {author} {\bibfnamefont {T.}~\bibnamefont
  {Nan}}, \bibinfo {author} {\bibfnamefont {C.~X.}\ \bibnamefont {Quintela}},
  \bibinfo {author} {\bibfnamefont {J.}~\bibnamefont {Irwin}}, \bibinfo
  {author} {\bibfnamefont {G.}~\bibnamefont {Gurung}}, \bibinfo {author}
  {\bibfnamefont {D.~F.}\ \bibnamefont {Shao}}, \bibinfo {author}
  {\bibfnamefont {J.}~\bibnamefont {Gibbons}}, \bibinfo {author} {\bibfnamefont
  {N.}~\bibnamefont {Campbell}}, \bibinfo {author} {\bibfnamefont
  {K.}~\bibnamefont {Song}}, \bibinfo {author} {\bibfnamefont {S.~Y.}\
  \bibnamefont {Choi}}, \bibinfo {author} {\bibfnamefont {L.}~\bibnamefont
  {Guo}}, \bibinfo {author} {\bibfnamefont {R.~D.}\ \bibnamefont {Johnson}},
  \bibinfo {author} {\bibfnamefont {P.}~\bibnamefont {Manuel}}, \bibinfo
  {author} {\bibfnamefont {R.~V.}\ \bibnamefont {Chopdekar}}, \bibinfo {author}
  {\bibfnamefont {I.}~\bibnamefont {Hallsteinsen}}, \bibinfo {author}
  {\bibfnamefont {T.}~\bibnamefont {Tybell}}, \bibinfo {author} {\bibfnamefont
  {P.~J.}\ \bibnamefont {Ryan}}, \bibinfo {author} {\bibfnamefont {J.~W.}\
  \bibnamefont {Kim}}, \bibinfo {author} {\bibfnamefont {Y.}~\bibnamefont
  {Choi}}, \bibinfo {author} {\bibfnamefont {P.~G.}\ \bibnamefont {Radaelli}},
  \bibinfo {author} {\bibfnamefont {D.~C.}\ \bibnamefont {Ralph}}, \bibinfo
  {author} {\bibfnamefont {E.~Y.}\ \bibnamefont {Tsymbal}}, \bibinfo {author}
  {\bibfnamefont {M.~S.}\ \bibnamefont {Rzchowski}},\ and\ \bibinfo {author}
  {\bibfnamefont {C.~B.}\ \bibnamefont {Eom}},\ }\bibfield  {title} {\bibinfo
  {title} {{Controlling spin current polarization through non-collinear
  antiferromagnetism}},\ }\href {https://doi.org/10.1038/s41467-020-17999-4}
  {\bibfield  {journal} {\bibinfo  {journal} {Nat. Commun.}\ }\textbf {\bibinfo
  {volume} {11}},\ \bibinfo {pages} {4671} (\bibinfo {year}
  {2020})}\BibitemShut {NoStop}%
\bibitem [{\citenamefont {Gurung}\ \emph {et~al.}(2020)\citenamefont {Gurung},
  \citenamefont {Shao},\ and\ \citenamefont {Tsymbal}}]{gurung2020}%
  \BibitemOpen
  \bibfield  {author} {\bibinfo {author} {\bibfnamefont {G.}~\bibnamefont
  {Gurung}}, \bibinfo {author} {\bibfnamefont {D.-F.}\ \bibnamefont {Shao}},\
  and\ \bibinfo {author} {\bibfnamefont {E.~Y.}\ \bibnamefont {Tsymbal}},\
  }\bibfield  {title} {\bibinfo {title} {{Spin-torque switching of noncollinear
  antiferromagnetic antiperovskites}},\ }\href
  {https://doi.org/10.1103/PhysRevB.101.140405} {\bibfield  {journal} {\bibinfo
   {journal} {Phys. Rev. B}\ }\textbf {\bibinfo {volume} {101}},\ \bibinfo
  {pages} {140405} (\bibinfo {year} {2020})}\BibitemShut {NoStop}%
\bibitem [{\citenamefont {Florez-Gomez}\ \emph {et~al.}(2022)\citenamefont
  {Florez-Gomez}, \citenamefont {Ibarra-Hernandez},\ and\ \citenamefont
  {Garcia-Castro}}]{florez-gomez2022}%
  \BibitemOpen
  \bibfield  {author} {\bibinfo {author} {\bibfnamefont {L.}~\bibnamefont
  {Florez-Gomez}}, \bibinfo {author} {\bibfnamefont {W.}~\bibnamefont
  {Ibarra-Hernandez}},\ and\ \bibinfo {author} {\bibfnamefont {A.~C.}\
  \bibnamefont {Garcia-Castro}},\ }\bibfield  {title} {\bibinfo {title}
  {Exploration of the spin-phonon coupling in the noncollinear
  antiferromagnetic antiperovskite mn\$\_3\${NiN}}\ }\href
  {https://doi.org/10.48550/ARXIV.2202.10544} {10.48550/ARXIV.2202.10544}
  (\bibinfo {year} {2022}),\ \bibinfo {note} {publisher: {arXiv} Version
  Number: 1}\BibitemShut {NoStop}%
\bibitem [{\citenamefont {Boldrin}\ \emph
  {et~al.}(2019{\natexlab{a}})\citenamefont {Boldrin}, \citenamefont
  {Samathrakis}, \citenamefont {Zemen}, \citenamefont {Mihai}, \citenamefont
  {Zou}, \citenamefont {Johnson}, \citenamefont {Esser}, \citenamefont
  {{McComb}}, \citenamefont {Petrov}, \citenamefont {Zhang},\ and\
  \citenamefont {Cohen}}]{boldrin2019}%
  \BibitemOpen
  \bibfield  {author} {\bibinfo {author} {\bibfnamefont {D.}~\bibnamefont
  {Boldrin}}, \bibinfo {author} {\bibfnamefont {I.}~\bibnamefont
  {Samathrakis}}, \bibinfo {author} {\bibfnamefont {J.}~\bibnamefont {Zemen}},
  \bibinfo {author} {\bibfnamefont {A.}~\bibnamefont {Mihai}}, \bibinfo
  {author} {\bibfnamefont {B.}~\bibnamefont {Zou}}, \bibinfo {author}
  {\bibfnamefont {F.}~\bibnamefont {Johnson}}, \bibinfo {author} {\bibfnamefont
  {B.~D.}\ \bibnamefont {Esser}}, \bibinfo {author} {\bibfnamefont {D.~W.}\
  \bibnamefont {{McComb}}}, \bibinfo {author} {\bibfnamefont {P.~K.}\
  \bibnamefont {Petrov}}, \bibinfo {author} {\bibfnamefont {H.}~\bibnamefont
  {Zhang}},\ and\ \bibinfo {author} {\bibfnamefont {L.~F.}\ \bibnamefont
  {Cohen}},\ }\bibfield  {title} {\bibinfo {title} {{Anomalous Hall effect in
  noncollinear antiferromagnetic Mn$_3$NiN thin films}},\ }\href
  {https://doi.org/10.1103/PhysRevMaterials.3.094409} {\bibfield  {journal}
  {\bibinfo  {journal} {Phys. Rev. Materials}\ }\textbf {\bibinfo {volume}
  {3}},\ \bibinfo {pages} {094409} (\bibinfo {year}
  {2019}{\natexlab{a}})}\BibitemShut {NoStop}%
\bibitem [{\citenamefont {You}\ \emph {et~al.}(2020)\citenamefont {You},
  \citenamefont {Bai}, \citenamefont {Chen}, \citenamefont {Zhou},
  \citenamefont {Zhou}, \citenamefont {Pan},\ and\ \citenamefont
  {Song}}]{you2020}%
  \BibitemOpen
  \bibfield  {author} {\bibinfo {author} {\bibfnamefont {Y.}~\bibnamefont
  {You}}, \bibinfo {author} {\bibfnamefont {H.}~\bibnamefont {Bai}}, \bibinfo
  {author} {\bibfnamefont {X.}~\bibnamefont {Chen}}, \bibinfo {author}
  {\bibfnamefont {Y.}~\bibnamefont {Zhou}}, \bibinfo {author} {\bibfnamefont
  {X.}~\bibnamefont {Zhou}}, \bibinfo {author} {\bibfnamefont {F.}~\bibnamefont
  {Pan}},\ and\ \bibinfo {author} {\bibfnamefont {C.}~\bibnamefont {Song}},\
  }\bibfield  {title} {\bibinfo {title} {{Room temperature anomalous Hall
  effect in antiferromagnetic Mn$_3$SnN films}},\ }\href
  {https://doi.org/10.1063/5.0032106} {\bibfield  {journal} {\bibinfo
  {journal} {Appl. Phys. Lett.}\ }\textbf {\bibinfo {volume} {117}},\ \bibinfo
  {pages} {222404} (\bibinfo {year} {2020})}\BibitemShut {NoStop}%
\bibitem [{\citenamefont {Hajiri}\ \emph {et~al.}(2019)\citenamefont {Hajiri},
  \citenamefont {Ishino}, \citenamefont {Matsuura},\ and\ \citenamefont
  {Asano}}]{hajiri2019}%
  \BibitemOpen
  \bibfield  {author} {\bibinfo {author} {\bibfnamefont {T.}~\bibnamefont
  {Hajiri}}, \bibinfo {author} {\bibfnamefont {S.}~\bibnamefont {Ishino}},
  \bibinfo {author} {\bibfnamefont {K.}~\bibnamefont {Matsuura}},\ and\
  \bibinfo {author} {\bibfnamefont {H.}~\bibnamefont {Asano}},\ }\bibfield
  {title} {\bibinfo {title} {{Electrical current switching of the noncollinear
  antiferromagnet Mn$_3$GaN}},\ }\href {https://doi.org/10.1063/1.5109317}
  {\bibfield  {journal} {\bibinfo  {journal} {Appl. Phys. Lett.}\ }\textbf
  {\bibinfo {volume} {115}},\ \bibinfo {pages} {052403} (\bibinfo {year}
  {2019})}\BibitemShut {NoStop}%
\bibitem [{\citenamefont {Boldrin}\ \emph {et~al.}(2018)\citenamefont
  {Boldrin}, \citenamefont {Mihai}, \citenamefont {Zou}, \citenamefont {Zemen},
  \citenamefont {Thompson}, \citenamefont {Ware}, \citenamefont {Neamtu},
  \citenamefont {Ghivelder}, \citenamefont {Esser}, \citenamefont {{McComb}},
  \citenamefont {Petrov},\ and\ \citenamefont {Cohen}}]{boldrin2018}%
  \BibitemOpen
  \bibfield  {author} {\bibinfo {author} {\bibfnamefont {D.}~\bibnamefont
  {Boldrin}}, \bibinfo {author} {\bibfnamefont {A.~P.}\ \bibnamefont {Mihai}},
  \bibinfo {author} {\bibfnamefont {B.}~\bibnamefont {Zou}}, \bibinfo {author}
  {\bibfnamefont {J.}~\bibnamefont {Zemen}}, \bibinfo {author} {\bibfnamefont
  {R.}~\bibnamefont {Thompson}}, \bibinfo {author} {\bibfnamefont
  {E.}~\bibnamefont {Ware}}, \bibinfo {author} {\bibfnamefont {B.~V.}\
  \bibnamefont {Neamtu}}, \bibinfo {author} {\bibfnamefont {L.}~\bibnamefont
  {Ghivelder}}, \bibinfo {author} {\bibfnamefont {B.}~\bibnamefont {Esser}},
  \bibinfo {author} {\bibfnamefont {D.~W.}\ \bibnamefont {{McComb}}}, \bibinfo
  {author} {\bibfnamefont {P.}~\bibnamefont {Petrov}},\ and\ \bibinfo {author}
  {\bibfnamefont {L.~F.}\ \bibnamefont {Cohen}},\ }\bibfield  {title} {\bibinfo
  {title} {{Giant Piezomagnetism in Mn$_3$NiN}},\ }\href
  {https://doi.org/10.1021/acsami.8b03112} {\bibfield  {journal} {\bibinfo
  {journal} {{ACS} Appl. Mater. Interfaces}\ }\textbf {\bibinfo {volume}
  {10}},\ \bibinfo {pages} {18863} (\bibinfo {year} {2018})}\BibitemShut
  {NoStop}%
\bibitem [{\citenamefont {Boldrin}\ \emph
  {et~al.}(2019{\natexlab{b}})\citenamefont {Boldrin}, \citenamefont {Johnson},
  \citenamefont {Thompson}, \citenamefont {Mihai}, \citenamefont {Zou},
  \citenamefont {Zemen}, \citenamefont {Griffiths}, \citenamefont {Gubeljak},
  \citenamefont {Ormandy}, \citenamefont {Manuel}, \citenamefont {Khalyavin},
  \citenamefont {Ouladdiaf}, \citenamefont {Qureshi}, \citenamefont {Petrov},
  \citenamefont {Branford},\ and\ \citenamefont {Cohen}}]{boldrin2019a}%
  \BibitemOpen
  \bibfield  {author} {\bibinfo {author} {\bibfnamefont {D.}~\bibnamefont
  {Boldrin}}, \bibinfo {author} {\bibfnamefont {F.}~\bibnamefont {Johnson}},
  \bibinfo {author} {\bibfnamefont {R.}~\bibnamefont {Thompson}}, \bibinfo
  {author} {\bibfnamefont {A.~P.}\ \bibnamefont {Mihai}}, \bibinfo {author}
  {\bibfnamefont {B.}~\bibnamefont {Zou}}, \bibinfo {author} {\bibfnamefont
  {J.}~\bibnamefont {Zemen}}, \bibinfo {author} {\bibfnamefont
  {J.}~\bibnamefont {Griffiths}}, \bibinfo {author} {\bibfnamefont
  {P.}~\bibnamefont {Gubeljak}}, \bibinfo {author} {\bibfnamefont {K.~L.}\
  \bibnamefont {Ormandy}}, \bibinfo {author} {\bibfnamefont {P.}~\bibnamefont
  {Manuel}}, \bibinfo {author} {\bibfnamefont {D.~D.}\ \bibnamefont
  {Khalyavin}}, \bibinfo {author} {\bibfnamefont {B.}~\bibnamefont
  {Ouladdiaf}}, \bibinfo {author} {\bibfnamefont {N.}~\bibnamefont {Qureshi}},
  \bibinfo {author} {\bibfnamefont {P.}~\bibnamefont {Petrov}}, \bibinfo
  {author} {\bibfnamefont {W.}~\bibnamefont {Branford}},\ and\ \bibinfo
  {author} {\bibfnamefont {L.~F.}\ \bibnamefont {Cohen}},\ }\bibfield  {title}
  {\bibinfo {title} {{The Biaxial Strain Dependence of Magnetic Order in Spin
  Frustrated Mn$_3$NiN Thin Films}},\ }\href
  {https://doi.org/10.1002/adfm.201902502} {\bibfield  {journal} {\bibinfo
  {journal} {Adv. Funct. Mater.}\ }\textbf {\bibinfo {volume} {29}},\ \bibinfo
  {pages} {1902502} (\bibinfo {year} {2019}{\natexlab{b}})}\BibitemShut
  {NoStop}%
\bibitem [{\citenamefont {Zemen}\ \emph {et~al.}(2017)\citenamefont {Zemen},
  \citenamefont {Gercsi},\ and\ \citenamefont {Sandeman}}]{zemen2017}%
  \BibitemOpen
  \bibfield  {author} {\bibinfo {author} {\bibfnamefont {J.}~\bibnamefont
  {Zemen}}, \bibinfo {author} {\bibfnamefont {Z.}~\bibnamefont {Gercsi}},\ and\
  \bibinfo {author} {\bibfnamefont {K.~G.}\ \bibnamefont {Sandeman}},\
  }\bibfield  {title} {\bibinfo {title} {{Piezomagnetism as a counterpart of
  the magnetovolume effect in magnetically frustrated Mn-based antiperovskite
  nitrides}},\ }\href {https://doi.org/10.1103/PhysRevB.96.024451} {\bibfield
  {journal} {\bibinfo  {journal} {Phys. Rev. B}\ }\textbf {\bibinfo {volume}
  {96}},\ \bibinfo {pages} {024451} (\bibinfo {year} {2017})}\BibitemShut
  {NoStop}%
\bibitem [{\citenamefont {Johnson}\ \emph {et~al.}(2021)\citenamefont
  {Johnson}, \citenamefont {Boldrin}, \citenamefont {Zemen}, \citenamefont
  {Pesquera}, \citenamefont {Kim}, \citenamefont {Moya}, \citenamefont {Zhang},
  \citenamefont {Singh}, \citenamefont {Samathrakis},\ and\ \citenamefont
  {Cohen}}]{johnson2021}%
  \BibitemOpen
  \bibfield  {author} {\bibinfo {author} {\bibfnamefont {F.}~\bibnamefont
  {Johnson}}, \bibinfo {author} {\bibfnamefont {D.}~\bibnamefont {Boldrin}},
  \bibinfo {author} {\bibfnamefont {J.}~\bibnamefont {Zemen}}, \bibinfo
  {author} {\bibfnamefont {D.}~\bibnamefont {Pesquera}}, \bibinfo {author}
  {\bibfnamefont {J.}~\bibnamefont {Kim}}, \bibinfo {author} {\bibfnamefont
  {X.}~\bibnamefont {Moya}}, \bibinfo {author} {\bibfnamefont {H.}~\bibnamefont
  {Zhang}}, \bibinfo {author} {\bibfnamefont {H.~K.}\ \bibnamefont {Singh}},
  \bibinfo {author} {\bibfnamefont {I.}~\bibnamefont {Samathrakis}},\ and\
  \bibinfo {author} {\bibfnamefont {L.~F.}\ \bibnamefont {Cohen}},\ }\bibfield
  {title} {\bibinfo {title} {{Strain dependence of Berry-phase-induced
  anomalous Hall effect in the non-collinear antiferromagnet Mn$_3$NiN}},\
  }\href {https://doi.org/10.1063/5.0072783} {\bibfield  {journal} {\bibinfo
  {journal} {Applied Physics Letters}\ }\textbf {\bibinfo {volume} {119}},\
  \bibinfo {pages} {222401} (\bibinfo {year} {2021})}\BibitemShut {NoStop}%
\bibitem [{\citenamefont {Ikhlas}\ \emph {et~al.}(2022)\citenamefont {Ikhlas},
  \citenamefont {Dasgupta}, \citenamefont {Theuss}, \citenamefont {Higo},
  \citenamefont {Kittaka}, \citenamefont {Ramshaw}, \citenamefont {Hicks},\
  and\ \citenamefont {Nakatsuji}}]{ikhlas2022}%
  \BibitemOpen
  \bibfield  {author} {\bibinfo {author} {\bibfnamefont {M.}~\bibnamefont
  {Ikhlas}}, \bibinfo {author} {\bibfnamefont {S.}~\bibnamefont {Dasgupta}},
  \bibinfo {author} {\bibfnamefont {F.}~\bibnamefont {Theuss}}, \bibinfo
  {author} {\bibfnamefont {T.}~\bibnamefont {Higo}}, \bibinfo {author}
  {\bibfnamefont {S.}~\bibnamefont {Kittaka}}, \bibinfo {author} {\bibfnamefont
  {B.~J.}\ \bibnamefont {Ramshaw}}, \bibinfo {author} {\bibfnamefont {C.~W.}\
  \bibnamefont {Hicks}},\ and\ \bibinfo {author} {\bibfnamefont
  {S.}~\bibnamefont {Nakatsuji}},\ }\bibfield  {title} {\bibinfo {title}
  {{Piezomagnetic switching of the anomalous Hall effect in an antiferromagnet
  at room temperature}},\ }\href {https://doi.org/10.1038/s41567-022-01645-5}
  {\bibfield  {journal} {\bibinfo  {journal} {Nat. Phys.}\ }\textbf {\bibinfo
  {volume} {18}},\ \bibinfo {pages} {1086–1093} (\bibinfo {year} {2022})},\
  \Eprint {https://arxiv.org/abs/https://doi.org/10.1038/s41567-022-01645-5}
  {https://doi.org/10.1038/s41567-022-01645-5} \BibitemShut {NoStop}%
\bibitem [{\citenamefont {You}\ \emph {et~al.}(2022)\citenamefont {You},
  \citenamefont {Lam}, \citenamefont {Wan}, \citenamefont {Wan}, \citenamefont
  {Zhu}, \citenamefont {Han}, \citenamefont {Bai}, \citenamefont {Zhou},
  \citenamefont {Qiao}, \citenamefont {Chen}, \citenamefont {Pan},
  \citenamefont {Liu},\ and\ \citenamefont {Song}}]{you2022}%
  \BibitemOpen
  \bibfield  {author} {\bibinfo {author} {\bibfnamefont {Y.}~\bibnamefont
  {You}}, \bibinfo {author} {\bibfnamefont {H.}~\bibnamefont {Lam}}, \bibinfo
  {author} {\bibfnamefont {C.}~\bibnamefont {Wan}}, \bibinfo {author}
  {\bibfnamefont {C.}~\bibnamefont {Wan}}, \bibinfo {author} {\bibfnamefont
  {W.}~\bibnamefont {Zhu}}, \bibinfo {author} {\bibfnamefont {L.}~\bibnamefont
  {Han}}, \bibinfo {author} {\bibfnamefont {H.}~\bibnamefont {Bai}}, \bibinfo
  {author} {\bibfnamefont {Y.}~\bibnamefont {Zhou}}, \bibinfo {author}
  {\bibfnamefont {L.}~\bibnamefont {Qiao}}, \bibinfo {author} {\bibfnamefont
  {T.}~\bibnamefont {Chen}}, \bibinfo {author} {\bibfnamefont {F.}~\bibnamefont
  {Pan}}, \bibinfo {author} {\bibfnamefont {J.}~\bibnamefont {Liu}},\ and\
  \bibinfo {author} {\bibfnamefont {C.}~\bibnamefont {Song}},\ }\bibfield
  {title} {\bibinfo {title} {{Anomalous Nernst Effect in an Antiperovskite
  Antiferromagnet}},\ }\href {https://doi.org/10.1103/PhysRevApplied.18.024007}
  {\bibfield  {journal} {\bibinfo  {journal} {Phys. Rev. Applied}\ }\textbf
  {\bibinfo {volume} {18}},\ \bibinfo {pages} {024007} (\bibinfo {year}
  {2022})}\BibitemShut {NoStop}%
\bibitem [{\citenamefont {{COMSOL
  Multiphysics\textsuperscript{\textregistered}}}()}]{comsol}%
  \BibitemOpen
  \bibfield  {author} {\bibinfo {author} {\bibnamefont {{COMSOL
  Multiphysics\textsuperscript{\textregistered}}}},\ }\href@noop {} {\bibinfo
  {title} {{v. 6.0. \url{www.comsol.com}. COMSOL AB, Stockholm,
  Sweden}}}\BibitemShut {NoStop}%
\bibitem [{\citenamefont {Park}\ \emph {et~al.}(2020)\citenamefont {Park},
  \citenamefont {Reichlova}, \citenamefont {Schlitz}, \citenamefont {Lammel},
  \citenamefont {Markou}, \citenamefont {Swekis}, \citenamefont {Ritzinger},
  \citenamefont {Kriegner}, \citenamefont {Noky}, \citenamefont {Gayles},
  \citenamefont {Sun}, \citenamefont {Felser}, \citenamefont {Nielsch},
  \citenamefont {Goennenwein},\ and\ \citenamefont {Thomas}}]{park2020}%
  \BibitemOpen
  \bibfield  {author} {\bibinfo {author} {\bibfnamefont {G.-H.}\ \bibnamefont
  {Park}}, \bibinfo {author} {\bibfnamefont {H.}~\bibnamefont {Reichlova}},
  \bibinfo {author} {\bibfnamefont {R.}~\bibnamefont {Schlitz}}, \bibinfo
  {author} {\bibfnamefont {M.}~\bibnamefont {Lammel}}, \bibinfo {author}
  {\bibfnamefont {A.}~\bibnamefont {Markou}}, \bibinfo {author} {\bibfnamefont
  {P.}~\bibnamefont {Swekis}}, \bibinfo {author} {\bibfnamefont
  {P.}~\bibnamefont {Ritzinger}}, \bibinfo {author} {\bibfnamefont
  {D.}~\bibnamefont {Kriegner}}, \bibinfo {author} {\bibfnamefont
  {J.}~\bibnamefont {Noky}}, \bibinfo {author} {\bibfnamefont {J.}~\bibnamefont
  {Gayles}}, \bibinfo {author} {\bibfnamefont {Y.}~\bibnamefont {Sun}},
  \bibinfo {author} {\bibfnamefont {C.}~\bibnamefont {Felser}}, \bibinfo
  {author} {\bibfnamefont {K.}~\bibnamefont {Nielsch}}, \bibinfo {author}
  {\bibfnamefont {S.~T.~B.}\ \bibnamefont {Goennenwein}},\ and\ \bibinfo
  {author} {\bibfnamefont {A.}~\bibnamefont {Thomas}},\ }\bibfield  {title}
  {\bibinfo {title} {{Thickness dependence of the anomalous Nernst effect and
  the Mott relation of Weyl semimetal Co$_2$MnGa thin films}},\ }\href
  {https://doi.org/10.1103/PhysRevB.101.060406} {\bibfield  {journal} {\bibinfo
   {journal} {Phys. Rev. B}\ }\textbf {\bibinfo {volume} {101}},\ \bibinfo
  {pages} {060406} (\bibinfo {year} {2020})}\BibitemShut {NoStop}%
\bibitem [{\citenamefont {Pu}\ \emph {et~al.}(2008)\citenamefont {Pu},
  \citenamefont {Chiba}, \citenamefont {Matsukura}, \citenamefont {Ohno},\ and\
  \citenamefont {Shi}}]{pu2008}%
  \BibitemOpen
  \bibfield  {author} {\bibinfo {author} {\bibfnamefont {Y.}~\bibnamefont
  {Pu}}, \bibinfo {author} {\bibfnamefont {D.}~\bibnamefont {Chiba}}, \bibinfo
  {author} {\bibfnamefont {F.}~\bibnamefont {Matsukura}}, \bibinfo {author}
  {\bibfnamefont {H.}~\bibnamefont {Ohno}},\ and\ \bibinfo {author}
  {\bibfnamefont {J.}~\bibnamefont {Shi}},\ }\bibfield  {title} {\bibinfo
  {title} {{Mott Relation for Anomalous Hall and Nernst Effects in
  ${\mathrm{Ga}}_{1\ensuremath{-}x}{\mathrm{Mn}}_{x}\mathrm{As}$ Ferromagnetic
  Semiconductors}},\ }\href {https://doi.org/10.1103/PhysRevLett.101.117208}
  {\bibfield  {journal} {\bibinfo  {journal} {Phys. Rev. Lett.}\ }\textbf
  {\bibinfo {volume} {101}},\ \bibinfo {pages} {117208} (\bibinfo {year}
  {2008})}\BibitemShut {NoStop}%
\bibitem [{\citenamefont {Boldrin}\ \emph
  {et~al.}(2019{\natexlab{c}})\citenamefont {Boldrin}, \citenamefont
  {Samathrakis}, \citenamefont {Zemen}, \citenamefont {Mihai}, \citenamefont
  {Zou}, \citenamefont {Johnson}, \citenamefont {Esser}, \citenamefont
  {McComb}, \citenamefont {Petrov}, \citenamefont {Zhang},\ and\ \citenamefont
  {Cohen}}]{Boldrin_AHE_Mn3NiN_Arxiv}%
  \BibitemOpen
  \bibfield  {author} {\bibinfo {author} {\bibfnamefont {D.}~\bibnamefont
  {Boldrin}}, \bibinfo {author} {\bibfnamefont {I.}~\bibnamefont
  {Samathrakis}}, \bibinfo {author} {\bibfnamefont {J.}~\bibnamefont {Zemen}},
  \bibinfo {author} {\bibfnamefont {A.}~\bibnamefont {Mihai}}, \bibinfo
  {author} {\bibfnamefont {B.}~\bibnamefont {Zou}}, \bibinfo {author}
  {\bibfnamefont {F.}~\bibnamefont {Johnson}}, \bibinfo {author} {\bibfnamefont
  {B.~D.}\ \bibnamefont {Esser}}, \bibinfo {author} {\bibfnamefont {D.~W.}\
  \bibnamefont {McComb}}, \bibinfo {author} {\bibfnamefont {P.~K.}\
  \bibnamefont {Petrov}}, \bibinfo {author} {\bibfnamefont {H.}~\bibnamefont
  {Zhang}},\ and\ \bibinfo {author} {\bibfnamefont {L.~F.}\ \bibnamefont
  {Cohen}},\ }\bibfield  {title} {\bibinfo {title} {{Anomalous Hall effect in
  noncollinear antiferromagnetic Mn$_3$NiN thin films}}\ }\href
  {https://doi.org/arXiv:1902.04357v1} {arXiv:1902.04357v1} (\bibinfo {year}
  {2019}{\natexlab{c}}),\ \bibinfo {note} {publisher: {arXiv} Version Number:
  1}\BibitemShut {NoStop}%
\bibitem [{\citenamefont {Smrcka}\ and\ \citenamefont
  {Streda}(1977)}]{Smrcka_1977}%
  \BibitemOpen
  \bibfield  {author} {\bibinfo {author} {\bibfnamefont {L.}~\bibnamefont
  {Smrcka}}\ and\ \bibinfo {author} {\bibfnamefont {P.}~\bibnamefont
  {Streda}},\ }\bibfield  {title} {\bibinfo {title} {{Transport coefficients in
  strong magnetic fields}},\ }\href
  {https://doi.org/10.1088/0022-3719/10/12/021} {\bibfield  {journal} {\bibinfo
   {journal} {J. Phys. C: Solid State Phys.}\ }\textbf {\bibinfo {volume}
  {10}},\ \bibinfo {pages} {2153} (\bibinfo {year} {1977})}\BibitemShut
  {NoStop}%
\bibitem [{\citenamefont {Dehkordi}\ \emph {et~al.}(2018)\citenamefont
  {Dehkordi}, \citenamefont {Bhattacharya}, \citenamefont {Darroudi},
  \citenamefont {Karakaya}, \citenamefont {Kucera}, \citenamefont {Ballato},
  \citenamefont {Adebisi}, \citenamefont {Gladden}, \citenamefont {Podila},
  \citenamefont {Rao}, \citenamefont {Alshareef},\ and\ \citenamefont
  {Tritt}}]{dehkordi2018}%
  \BibitemOpen
  \bibfield  {author} {\bibinfo {author} {\bibfnamefont {A.~M.}\ \bibnamefont
  {Dehkordi}}, \bibinfo {author} {\bibfnamefont {S.}~\bibnamefont
  {Bhattacharya}}, \bibinfo {author} {\bibfnamefont {T.}~\bibnamefont
  {Darroudi}}, \bibinfo {author} {\bibfnamefont {M.}~\bibnamefont {Karakaya}},
  \bibinfo {author} {\bibfnamefont {C.}~\bibnamefont {Kucera}}, \bibinfo
  {author} {\bibfnamefont {J.}~\bibnamefont {Ballato}}, \bibinfo {author}
  {\bibfnamefont {R.}~\bibnamefont {Adebisi}}, \bibinfo {author} {\bibfnamefont
  {J.~R.}\ \bibnamefont {Gladden}}, \bibinfo {author} {\bibfnamefont
  {R.}~\bibnamefont {Podila}}, \bibinfo {author} {\bibfnamefont {A.~M.}\
  \bibnamefont {Rao}}, \bibinfo {author} {\bibfnamefont {H.~N.}\ \bibnamefont
  {Alshareef}},\ and\ \bibinfo {author} {\bibfnamefont {T.~M.}\ \bibnamefont
  {Tritt}},\ }\bibfield  {title} {\bibinfo {title} {{Optimizing thermal
  conduction in bulk polycrystalline SrTiO$_{3-\delta}$ ceramics via oxygen
  non-stoichiometry}},\ }\href {https://doi.org/10.1557/mrc.2018.220}
  {\bibfield  {journal} {\bibinfo  {journal} {MRS Communications}\ }\textbf
  {\bibinfo {volume} {8}},\ \bibinfo {pages} {1470} (\bibinfo {year}
  {2018})}\BibitemShut {NoStop}%
\bibitem [{\citenamefont {Schreier}\ \emph {et~al.}(2013)\citenamefont
  {Schreier}, \citenamefont {Roschewsky}, \citenamefont {Dobler}, \citenamefont
  {Meyer}, \citenamefont {Huebl}, \citenamefont {Gross},\ and\ \citenamefont
  {Goennenwein}}]{schreier2013}%
  \BibitemOpen
  \bibfield  {author} {\bibinfo {author} {\bibfnamefont {M.}~\bibnamefont
  {Schreier}}, \bibinfo {author} {\bibfnamefont {N.}~\bibnamefont
  {Roschewsky}}, \bibinfo {author} {\bibfnamefont {E.}~\bibnamefont {Dobler}},
  \bibinfo {author} {\bibfnamefont {S.}~\bibnamefont {Meyer}}, \bibinfo
  {author} {\bibfnamefont {H.}~\bibnamefont {Huebl}}, \bibinfo {author}
  {\bibfnamefont {R.}~\bibnamefont {Gross}},\ and\ \bibinfo {author}
  {\bibfnamefont {S.~T.~B.}\ \bibnamefont {Goennenwein}},\ }\bibfield  {title}
  {\bibinfo {title} {{Current heating induced spin Seebeck effect}},\ }\href
  {https://doi.org/10.1063/1.4839395} {\bibfield  {journal} {\bibinfo
  {journal} {Appl. Phys. Lett.}\ }\textbf {\bibinfo {volume} {103}},\ \bibinfo
  {pages} {242404} (\bibinfo {year} {2013})}\BibitemShut {NoStop}%
\end{thebibliography}
\end{document}